\documentclass[aps,superscriptaddress,showpacs,longbibliography,floatfix]{revtex4-2}

\usepackage[utf8]{inputenc}
\usepackage{slashed}
\pdfoutput=1

\usepackage{color}
\usepackage{graphicx}   
\usepackage{bm}
\usepackage{amsmath}
\usepackage{amsfonts}
\usepackage{hyperref}

\def\be{\begin{equation}}
\def\ee{\end{equation}}
\def\ba{\begin{eqnarray}}
\def\ea{\end{eqnarray}}

\begin{document}

\title{Phase Diagram and Finite Temperature Properties of Negative Coupling Scalar Field Theory}

\author{Paul Romatschke}
\affiliation{Institut für Theoretische Physik, TU Wien, Wiedner Hauptstraße 8-10, 1040 Wien, Austria}

\begin{abstract}
In this work, I consider scalar field theory with negative quartic self-interaction, corresponding to an upside-down classical potential. Despite not possessing a classically stable ground state, such potentials are known to behave properly when treated quantum mechanically, leading to stable and unitary time evolution. Using two different saddle-point expansions for the same theory, I discuss the phase diagram in terms of bare parameters in Euclidean dimensions one to four, as well as the generalization to finite temperature. Comparing to other methods where available, I find that negative coupling field theory is a promising candidate for an interacting scalar field theory in the continuum. In particular, in four dimensions it exploits a loophole in mathematical proofs of quantum triviality, suggesting that negative coupling scalar field theory could offer a UV-complete and interacting description of the Higgs.
\end{abstract}

\maketitle

\section{Introduction}

In classical physics, upside-down potentials do not have stable ground states, as those who choose to play golf on a mountain will readily acknowledge. In quantum field theory, the notion that all ``good'' potentials must have stable classical ground states is enshrined as an ``obvious'' condition in almost all textbooks, and early ideas to build quantum field theories with negative coupling \cite{Symanzik:1973hx} were derided as ``nonsense'' by some of the most distinguished members of the community \cite{Coleman:1973sx}.

The same ``obvious'' condition could be applied to the electron in the hydrogen atom, because a charged classical particle on a circular orbit continually emits electromagnetic radiation, leading to orbital decay, thus preventing a stable configuration. Yet the electronic ground state of the hydrogen atom is quantum mechanically stable, despite possessing no classically stable ground state.

Further information surfaced in the twilight of the last millennium \cite{Bender:1998ke}, where it was found that even non-Hermitian potentials can lead to real and bounded eigenspectra of quantum mechanical Hamiltonians if they possess certain characteristics \cite{bender1999,Dorey:2001uw,Bender:2007nj,Dorey:2007zx}. Using the similarities between Schr\"odinger's wave equation and wave optics and open quantum systems \cite{2018NatPh..14...11E}, experimental verification of these ``exotic'' theory predictions was possible (cf. Ref.~\cite{2010NatPh...6..192R}), creating a new subfield of material physics, see for instance Ref.~\cite{2019PhRvL.122u3201D}.

Unlike negative coupling quantum mechanics, very little is known about negative coupling quantum field theory. Studies of non-classical quantum field theory were first conducted for $i\phi^3$ potentials in Refs.~\cite{PhysRevLett.40.1610,PhysRevLett.54.1354,Li:2024xms,ArguelloCruz:2025zuq}, and more recently for a broader class of complex potentials in Refs.~\cite{Bender:2018pbv,Felski:2021evi,Branchina:2021czr}. Subsequently, non-Hermitian quantum field theory has been considered for fermionic theories (cf. Refs.~\cite{Beygi:2019qab,Felski:2020vrm,Mavromatos:2020hfy,Felski:2021bdg,Mavromatos:2021hpe}), quantum gravity (cf. Refs.~\cite{Mavromatos:2024ozk,Kuntz:2024rzu}) and lattice systems (cf. Refs.~\cite{Romatschke:2023fax,Ogilvie:2024vde}). Studies of negative coupling quantum field theory are even more recent, see e.g. Refs.~\cite{Ai:2022csx,Romatschke:2022jqg,Lawrence:2023woz,Weller:2023jhc,Chen:2024ynx,Romatschke:2024cld,Barberena:2025ibo}, but potentially important as descriptions of the Higgs field, cf. Refs.~\cite{Fring:2019hue,Fring:2020bvr,Romatschke:2024hpb}.

One of the issues that studies of negative coupling field theory have faced is that established techniques such as weak coupling perturbation theory and lattice field theory using Monte Carlo importance sampling fail, simply because the classical probabilistic interpretation underlying these established techniques is absent. By contrast, non-perturbative methods such as large N expansions pioneered in the 1970s \cite{Abbott:1975bn,Linde:1976qh} were found to work well \cite{Romatschke:2022jqg,Romatschke:2023sce,Romatschke:2023ztk,Weller:2023jhc,Romatschke:2024yhx} whereas contour-deformation methods for lattice studies \cite{Lawrence:2023woz,Romatschke:2023fax,Lawrence:2022afv} work in principle, but are numerically too expensive for extracting continuum physics in space-time dimensions $d>2$.

In this work, the focus is on uncovering the broad qualitative features of the phase diagram of negative coupling scalar field theory, both at zero and finite temperature. The main tool for this study is the use of two distinct saddle-point expansions of the theory, which were presented and cross-checked against other methods for the case of positive-coupling field theory in Ref.~\cite{Romatschke:2026tam}. It should be emphasized that while the expansions are systematically improvable, there is no small expansion parameter, and therefore truncations correspond to uncontrolled approximations of the theory. However, the truncations made in this study have the advantage that results are analytically tractable, and can be compared to other non-perturbative methods whenever available (to date basically only quantum mechanics, cf. Ref.~\cite{Jones:2006qs}) to check on the quality of the approximation.

In this sense, the present work is intended as a survey of the phase diagram of negative coupling field theory both at zero and finite temperature, with the aim of identifying the approximate location of features such as transition lines that may subsequently be probed by numerical methods, such as lattice field theory \cite{Romatschke:2023fax}.

\section{Calculation}

I consider a scalar field $\phi$ with Euclidean action
\be
\label{action}
S=\int dx \left[\frac{1}{2}\partial_\mu \phi \partial_\mu \phi+\frac{1}{2}m_B^2 \phi^2-g_B \phi^4\right]\,,
\ee
where the time-like Euclidean direction is a circle with radius of inverse temperature, $\beta=\frac{1}{T}$. For further simplicity, it will be useful to state the results for the propagator and pressure of a free massive boson in dimension $d$ at finite temperature:
\be
\Delta_{\rm free}(M,T)=T \sum_n \int \frac{d^{d-1}k}{(2\pi)^d} \frac{1}{\omega_n^2+k^2+M^2}\,,\quad p_{\rm free}(M,T)=-\frac{T}{2}\sum_n \int \frac{d^{d-1}k}{(2\pi)^d} \ln\left[{\omega_n^2+k^2+M^2}\right]\,,
\ee
where $\omega_n=2 \pi n T$ are the bosonic Matsubara frequencies. Some integrations may be performed to give the results
\ba
\Delta_{\rm free}(M,T)&=&\frac{\Gamma\left(1-\frac{d}{2}\right)}{(4\pi)^{\frac{d}{2}}}M^{d-2}+\int \frac{d^{d-1}k}{(2\pi)^{d-1}} \frac{n_B\left(\sqrt{k^2+M^2}\right)}{\sqrt{k^2+M^2}}\,,\\
p_{\rm free}(M,T)&=&\frac{\Gamma\left(-\frac{d}{2}\right)}{2 (4\pi)^{\frac{d}{2}}} M^d-T\int \frac{d^{d-1}k}{(2\pi)^{d-1}}\ln\left(1-e^{-\beta \sqrt{k^2+M^2}}\right)\,,
\ea
where $n_B(x)=\frac{1}{e^{\beta x}-1}$ is the Bose-Einstein distribution function. Yet another form can be given in terms of 
the high-temperature expansion \cite[Eq. (2.90)]{Laine:2016hma}:
\ba
 \label{htep}
  \Delta_{\rm free}(M,T)&=&\frac{T M^{d-3}\Gamma\left(\frac{3-d}{2}\right)}{(4\pi)^{\frac{d-1}{2}}}+\frac{2 T}{(4\pi)^{\frac{d-1}{2}}(2 \pi T)^{3-d}}\sum_{l=0}^{\infty}\left[\frac{-M^2}{(2\pi T)^2}\right]^l
  \frac{\Gamma\left(l+\frac{3-d}{2}\right)}{\Gamma\left(l+1\right)} \zeta\left(2 l+3-d\right)\,,\\
  p_{\rm free}(M,T)&=&\frac{T^d\zeta(d)\Gamma\left(\frac{d}{2}\right)}{\pi^{\frac{d}{2}}}-\frac{ T M^{d-1}\Gamma\left(\frac{3-d}{2}\right)}{(d-1)(4\pi)^{\frac{d-1}{2}}}+\frac{T}{(4\pi)^{\frac{d-1}{2}}(2 \pi T)^{1-d}}\sum_{l=0}^{\infty}\left[\frac{-M^2}{(2\pi T)^2}\right]^{l+1}
  \frac{\Gamma\left(l+\frac{3-d}{2}\right)}{\Gamma\left(l+2\right)} \zeta\left(2 l+3-d\right)\,,\nonumber
  \ea
  where $p_{SB}(T)=\frac{T^d\zeta(d)\Gamma\left(\frac{d}{2}\right)}{\pi^{\frac{d}{2}}}$ can be recognized as the Stefan-Boltzmann pressure for a free massless boson in $d$ dimension.

\subsection{Symmetric Saddle Expansion}

Using the mathematical identity
\be
e^{g\phi^4}=\int \frac{d\zeta}{\sqrt{16 g \pi}} e^{-\frac{\zeta^2}{16g}+\frac{1}{2}\zeta \phi^2}\,,
\ee
I rewrite the action of the theory into a form that only contains $\phi$ quadratically. Using the R1 resummation \cite{Romatschke:2019rjk}, I find for the partition function of the theory
\be
Z=\int d\zeta_0 e^{\beta V\left(p_{\rm free}(M,T)-\frac{\zeta_0^2}{16 g_B}-2 g_B \Delta_{\rm free}^2(M,T)\right)}\,,
\ee
where $\beta V$ is the volume of space-time, and
\be
M^2\equiv m_B^2+\zeta_0+\nu^2, \quad \nu^2=-8 g_B \Delta_{\rm free}(M,T)\,,
\ee
is the pole-mass of the field $\phi$.
In the large volume limit, the partition function may be calculated using the saddle point method, finding
\be
\frac{\ln Z}{\beta V}\equiv p(M,T)=p_{\rm free}(M,T)-\frac{\zeta_0^2}{16 g_B}-2 g_B \Delta^2_{\rm free}(M,T)\,,
\ee
with the saddle point condition 
\be
0=\frac{\partial p(M,T)}{\partial M^2}=-\frac{1}{2} \Delta_{\rm free}(M,T)-\frac{\zeta_0}{8g_B}\,,
\ee
This implies $\zeta_0=-4 g_B \Delta_{\rm free}(M,T)$, such that $\nu^2=2 \zeta_0$ and the saddle point condition can be expressed in terms of the pole mass \cite{Romatschke:2019rjk}. One finds for the symmetric phase 
\be
M^2=m_B^2-12 g_B \Delta_{\rm free}(M,T)\,, \quad p(M,T)=p_{\rm free}(M,T)-\frac{(M^2-m_B^2)^2}{48 g_B}\,.
\ee

\subsection{Broken Phase Saddle Expansion}

In contrast to the symmetric phase expansion, one can expand the action using
\be
\phi(x)=\phi_0+\xi(x)\,,
\ee
corresponding to an expansion around a phase with $\langle \phi \rangle \neq 0$. Using the R1 resummation \cite{Romatschke:2026tam}, one obtains the pressure in the broken phase as
\be
\tilde p(\tilde M,T)=-\frac{m_B^2 \phi_0^2}{2}+g_B \phi_0^4+p_{\rm free}(\tilde M,T)-3 g_B \Delta_{\rm free}^2(\tilde M,T)\,,
\ee
where the pole mass $\tilde M$ of the fluctuation field $\xi$ fulfills
\be
\tilde M^2=m_B^2-12 g_B \phi_0^2+\tilde \nu^2\,,\quad \tilde \nu^2=-12 g_B \Delta_{\rm free}(\tilde M, T)\,.
\ee
In the large volume limit, the value of $\phi_0$ is fixed through the saddle point condition
\be
0=\frac{\partial \tilde p(\tilde M,T)}{\partial \phi_0^2}=-\frac{m_B^2}{2}+2 g_B \phi_0^2+6 g_B\Delta_{\rm free}(\tilde M,T)\,.
\ee
Using this condition, the pole mass can be written as
\be
\tilde M^2=-2 m_B^2+24 g_B \Delta_{\rm free}(\tilde M,T)\,,
\ee
so that the saddle point condition and broken phase pressure simplify to
\be
\phi_0^2=-\frac{\tilde M^2}{8g_B}\,,\quad \tilde p(\tilde M,T)=p_{\rm free}(\tilde M,T)+\frac{\tilde M^4}{96g_B}+\frac{\tilde M^2 m_B^2}{24 g_B}-\frac{m_B^4}{48g_B}\,.
\ee
  
\section{d=1: ${\cal PT}$-Symmetric Quantum Mechanics}

For the case of quantum mechanics d=1, the symmetric phase and broken phase results simplify to
\ba
\label{d1res}
p(M,T)=-\frac{M}{2}-T \ln\left(1-e^{-\beta M}\right)-\frac{(M^2-m_B^2)^2}{48 g_B}\,,\quad M^2=m_B^2-\frac{6 g_B}{M}-\frac{12 g_B n_B(M)}{M}\,,\\
\tilde p(\tilde M,T)=-\frac{\tilde M}{2}-T \ln\left(1-e^{-\beta \tilde M}\right)+\frac{\tilde M^4}{96g_B}+\frac{\tilde M^2 m_B^2}{24 g_B}-\frac{m_B^4}{48g_B}\,,\quad \tilde M^2=-2 m_B^2+\frac{12 g_B}{\tilde M}+\frac{24 g_B n_B(\tilde M)}{\tilde M} \,.\nonumber
\ea

\begin{figure}
  \includegraphics[width=.48\linewidth]{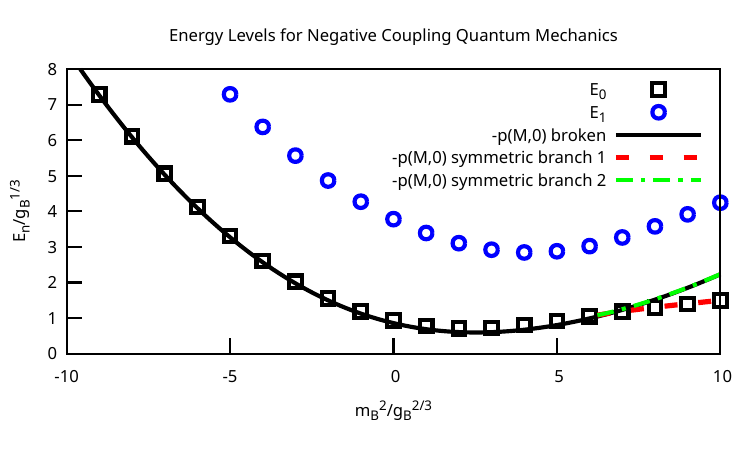}\hfill
  \includegraphics[width=.48\linewidth]{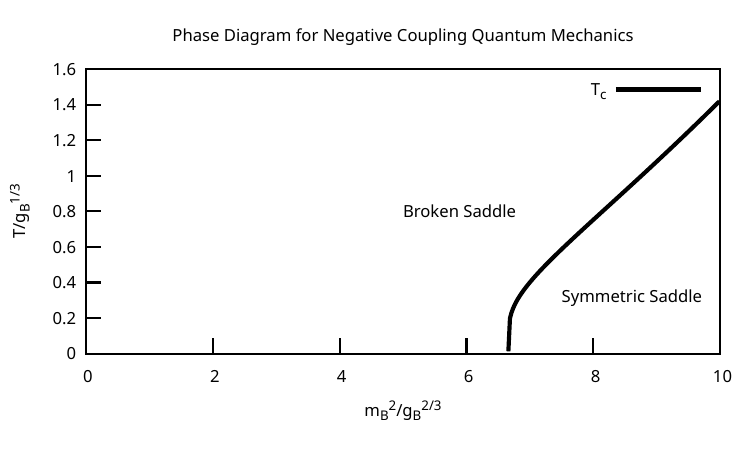}
  \caption{\label{fig:1dt0} Left: Comparison of lowest lying eigenvalues $E_0,E_1$ of the Hamiltonian (\ref{hami}) to minus the pressure $p(M,T=0),\tilde{p}(\tilde{M},T=0)$ from the symmetric and broken saddles obtained in the R1-level resummation. Right: Phase Diagram at finite temperature, indicating which saddle is thermodynamically preferred. Note that there is no actual phase transition here, just a transition from one saddle to another saddle. See text for details.}
\end{figure}

Let me first discuss the case of zero temperature $T=0$, where one finds solutions $M\in \mathbb{R}^+$ for $m_B^2\geq 243^{\frac{1}{3}} g_B^{\frac{2}{3}}\simeq 6.24 g_B^{\frac{2}{3}}$ for the symmetric phase, whereas solutions $\tilde M\in \mathbb R^+$ exist for all $m_B^2$. Using the criterion outlined in \cite{Romatschke:2026tam}, these solutions correspond to different phases of the theory, with the phase with the highest pressure (lowest free energy) being thermodynamically favored. Comparing $p(M,0)-\tilde p(\tilde M,0)$ for the solutions outlined above, one finds that there is a transition from a broken phase for small $\frac{m_B^2}{g_B^{\frac{2}{3}}}$ to a symmetric phase at
\be
m_B^2\simeq 6.66 g_B^{\frac{2}{3}}\,.
\ee
As discussed in the supplemental material of Ref.~\cite{Romatschke:2026tam}, a transition from broken to symmetric phase is not taking place in the quantum mechanical theory, even though the pressure in the symmetric phase does give an accurate numerical description of the ground state energy of the system. This is verified in Fig.~\ref{fig:1dt0} through the numerical diagonalization of the Hamiltonian
\be
\label{hami}
{\cal H}=\frac{p^2}{2}+4 g_B x^4+\sqrt{2 g_B} x-x^2 m_B^2+\frac{m_B^4}{16g_B}\,,
\ee
which is isospectral to the negative coupling (${\cal PT}$-symmetric) Hamiltonian ${\cal H}=\frac{p^2}{2}+\frac{1}{2}m_B^2 x^2-g_B x^4$ \cite{Jones:2006qs,Bender:2007nj}.

At finite temperature, real-valued solutions for the symmetric pole mass $M$ exist only below $T<T_c(m_B^2)$, whereas real-valued solutions exist for $\tilde{M}$ for all $m_B^2,T$. It is possible to track the solution with the highest pressure for all $m_B^2,T$, and one obtains the phase diagram for the preferred thermodynamic saddle shown in Fig.~\ref{fig:1dt0}. In the high temperature limit $T\gg m_B$, results (\ref{d1res}) simplify to
\ba
p(M,T)=-\frac{M}{2}-T \ln\left(\beta M\right)-\frac{M^4}{48 g_B}\,,\quad M^2=-\frac{12 g_B T}{M^2}\,,\\
\tilde p(\tilde M,T)=-\frac{\tilde M}{2}-T \ln\left(\beta \tilde M\right)+\frac{\tilde M^4}{96g_B}\,,\quad \tilde M^2=\frac{24 g_B T}{\tilde M^2} \,.\nonumber
\ea
In the case of the symmetric saddle, there are no real-valued solutions $M$ on the principal Riemann sheet, but instead one has 
\be
M^2=\pm i (12 g_B T)^{\frac{1}{4}}\,,
\label{2dcomplex}
\ee
and complex-valued pressure at high temperature. This matches the well-documented cases of complex-valued pressure at high temperature for symmetric saddle expansions in the literature in various dimensions, cf. the discussions in Ref.~\cite{Ai:2022csx,Romatschke:2022jqg,Kamata:2023opn,Lawrence:2023woz,Kamata:2024tyb,Romatschke:2024cld}. Ref.~\cite{Romatschke:2024cld} proposed a resolution of the problem by using solutions to the saddle-point condition on non-principal Riemann sheets.

In this work, I point out a different way from Ref.~\cite{Romatschke:2024cld} that also resolves the issue of complex valued pressures at high temperature. In particular, the broken saddle point condition has real-valued solutions
\be
\tilde M=(24 g_B T)^{\frac{1}{4}}\,,
\ee
on the principal Riemann sheet, and the resulting pressure in the high temperature limit is
\be
\tilde p(\tilde M\ll T)=-\frac{T}{4} \ln\left(\frac{24 g_B}{e^1 T^3}\right)+\ldots\,,
\ee
which is real-valued and positive for high-temperatures, as advertised.

\section{d=2: Negative Coupling Field Theory}

For the case of $d=2-2\varepsilon$, the symmetric phase and broken phase results lead to
\ba
\label{d2res}
p(M,T)&=&-\frac{M^2}{8\pi}\left(\frac{1}{\varepsilon}+\ln\frac{\bar\mu^2 e^1}{M^2}\right)-\frac{(M^2-m_B^2)^2}{48 g_B}+\frac{T M}{\pi}\sum_{n=1}^\infty \frac{K_1\left(n \beta M\right)}{n}\,,\\
M^2&=&m_B^2-\frac{3 g_B}{\pi}\left(\frac{1}{\varepsilon}+\ln\frac{\bar\mu^2}{M^2}\right)-\frac{12 g_B}{\pi}\sum_{n=1}^\infty K_0\left(n \beta M\right)\,,\nonumber\\
\tilde p(\tilde M,T)&=&-\frac{\tilde M^2}{8\pi}\left(\frac{1}{\varepsilon}+\ln\frac{\bar\mu^2 e^1}{\tilde M^2}\right)+\frac{\tilde M^4}{96g_B}+\frac{\tilde M^2m_B^2}{24 g_B}-\frac{m_B^4}{48 g_B}+\frac{T \tilde M}{\pi}\sum_{n=1}^\infty \frac{K_1\left(n \beta \tilde M\right)}{n}\,,\\
\tilde M^2&=&-2 m_B^2+\frac{6 g_B}{\pi}\left(\frac{1}{\varepsilon}+\ln\frac{\bar\mu^2}{\tilde M^2}\right)+\frac{24 g_B}{\pi}\sum_{n=1}^\infty K_0\left(n \beta \tilde M\right)\,,\nonumber
\ea
where $\bar\mu^2=4\pi \mu^2 e^{-\gamma_E}$ is the $\overline{\rm MS}$ renormalization scale and I used
\be
\int \frac{dk}{2\pi}\ln\left(1-e^{-\beta \sqrt{k^2+M^2}}\right)=-\frac{M}{\pi}\sum_{n=1}^\infty \frac{K_1\left(n \beta M\right)}{n}\,.
\ee

The divergencies can be non-perturbatively renormalized using
\be
m_B^2=m_R^2(\bar\mu)+\frac{3 g_B}{\pi \varepsilon}\,,
\ee
which leads to the running renormalized mass in the $\overline{\rm MS}$ scheme
\be
m_R^2(\bar\mu)=\frac{3 g_B}{\pi}\ln\frac{\bar\mu^2}{\Lambda_{\overline{\rm MS}}^2}\,,
\ee
where $\Lambda_{\overline{\rm MS}}$ is the $\overline{\rm MS}$ parameter. Plugging this result for $m_B^2$ into the above formulas leads to
\ba
\label{d2res}
p(M,T)=-\frac{M^2}{8\pi}\ln\frac{\Lambda_{\overline{\rm MS}}^2 e^1}{M^2}-\frac{M^4+m_B^4}{48 g_B}+\sum_{n=1}^\infty \frac{T M K_1\left(n \beta M\right)}{\pi n}\,,
-\frac{\pi M^2}{3 g_B}=\ln\frac{\Lambda_{\overline{\rm MS}}^2}{M^2}+4 \sum_{n=1}^\infty K_0\left(n \beta M\right)\,,\nonumber\\
\tilde p(\tilde M,T)=-\frac{\tilde M^2}{8\pi}\ln\frac{\Lambda_{\overline{\rm MS}}^2 e^1}{\tilde M^2}+\frac{\tilde M^4-2 m_B^4}{96g_B}+\sum_{n=1}^\infty \frac{T \tilde M K_1\left(n \beta \tilde M\right)}{\pi n}\,,\quad
\frac{\pi \tilde M^2}{6 g_B}=\ln\frac{\Lambda_{\overline{\rm MS}}^2}{\tilde M^2}+4\sum_{n=1}^\infty K_0\left(n \beta \tilde M\right)\,.\nonumber
\ea

\begin{figure}
  \includegraphics[width=.48\linewidth]{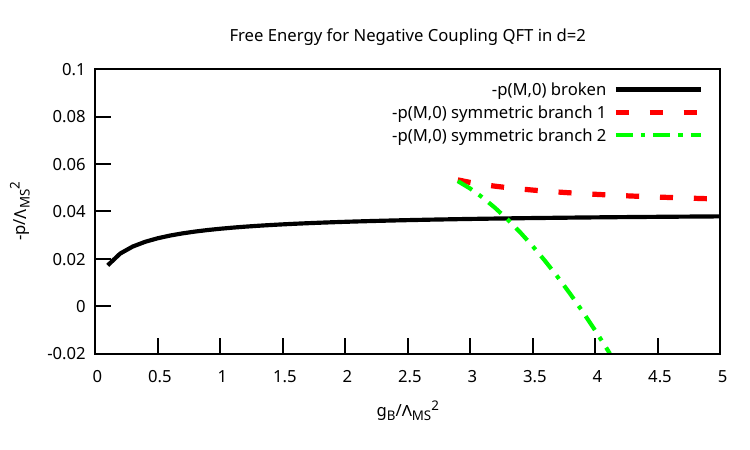}
  \hfill
  \includegraphics[width=.48\linewidth]{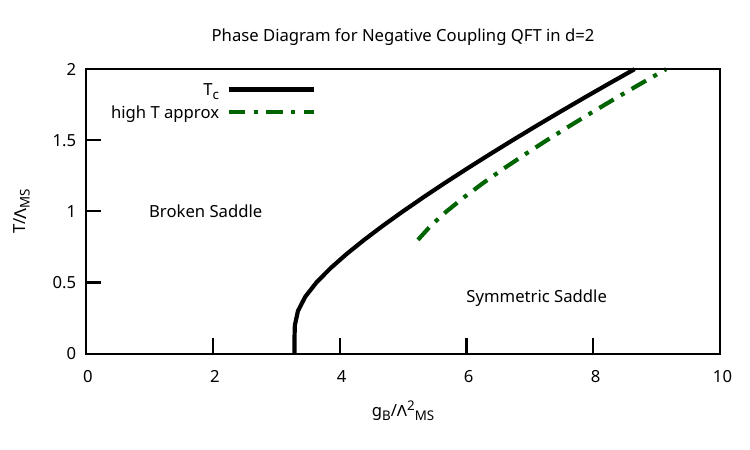}
  \caption{\label{fig:2d} Left: Free Energy (negative pressure) from the symmetric and broken phase saddle expansions, as a function of the coupling. Right: Phase diagram at finite temperature, indicating which saddle is thermodynamically preferred. Line labeled 'high T approx' is Eq.~(\ref{d2hightap}). See text for details.}
\end{figure}

At zero temperature $T=0$, a pair of real-valued solutions for the symmetric phase exist for $g_B\geq \frac{\pi e^1}{3}\Lambda_{\overline{\rm MS}}^2$ and are given by
\be
M^2=-\frac{3 g_B}{\pi}W_{-1,0}\left(-\frac{\pi \Lambda_{\overline{\rm MS}}^2}{3 g_B}\right)\,,
\label{d2t0gap}
\ee
where $W_k$ denotes the Lambert W-function of branch $k$. For the broken phase, real-valued solutions exist for all values of $g_B\geq 0$ and are given by
\be
\label{d2t0tildegap}
\tilde M^2=\frac{6 g_B}{\pi}W_{0}\left(\frac{\pi \Lambda_{\overline{\rm MS}}^2}{6 g_B}\right)\,.
\ee
While $p,\tilde p$ are divergent, the difference
\be
\Delta p(M,\tilde M,T)\equiv p(M,T)-\tilde p(\tilde M,T)\,,
\ee
is finite, and can be calculated to decide which saddle offers the thermodynamically preferred (higher pressure) phase. At zero temperature, one finds that starting at small $g_B$, the dominant phase is given by the broken phase until a critical value of
\be
g_c\equiv \frac{g_B}{\Lambda_{\overline{\rm MS}}^2}\simeq 3.2894\,,
\ee
above which the dominant phase is given by the symmetric phase with solution branch $k=-1$ in (\ref{d2t0gap}). I note that at the critical coupling, the symmetric pole mass fulfills
\be
\frac{g_B}{M^2}=-\frac{\pi}{3 W_{-1}\left(-\frac{\pi}{3 g_c}\right)}\simeq 0.64\,, 
\ee
which lends itself to comparison to lattice-based approaches (see e.g. Ref.~\cite{Schaich:2009jk} for the positive coupling situation). A plot of the free energies at $T=0$ for the different phases is shown in Fig.~\ref{fig:2d}.

\subsection{High temperature and dimensional reduction} 

As temperature $T$ is increased, the critical coupling value where $\Delta p(M,\tilde M,T)=0$ increases as well (see Fig.\ref{fig:2d}). In the high temperature limit $T\gg \Lambda_{\overline{\rm MS}}$, one can use the expansions (\ref{htep}) to express (\ref{d2res}) after renormalization as
\ba
\label{d2highT}
p(M,T)=\frac{T^2\pi}{6}-\frac{T M}{2}-\frac{M^2}{8\pi}\ln\frac{\Lambda_{\overline{\rm MS}}^2 e^{2 \gamma_E}}{(4 \pi T)^2}-\frac{M^4+m_B^4}{48g_B}\,,\quad -\frac{\pi M^2}{3 g_B}=\frac{2 \pi T}{M}+\ln\frac{\Lambda_{\overline{\rm MS}}^2 e^{2 \gamma_E}}{(4 \pi T)^2}\,,\\
\tilde p(\tilde M,T)=\frac{T^2\pi}{6}-\frac{T \tilde M}{2}-\frac{\tilde M^2}{8\pi}\ln\frac{\Lambda_{\overline{\rm MS}}^2 e^{2 \gamma_E}}{(4 \pi T)^2}+\frac{\tilde M^4-2m_B^4}{96 g_B}\,,\quad \frac{\pi \tilde M^2}{6 g_B}=\frac{2 \pi T}{\tilde M}+\ln\frac{\Lambda_{\overline{\rm MS}}^2 e^{2 \gamma_E}}{(4 \pi T)^2}\,.
\ea

For high temperatures, the action (\ref{action}) dimensionally reduces to
\be
\label{dimred}
S_{\rm red}= \int d^{d-1}x  \left[\frac{1}{2}\partial_i \phi \partial_i \phi+\frac{1}{2}m_{B,(d-1)}^2 \phi^2-T g_{B,(d)} \phi^4\right]\,,
\ee
where $i=1,\ldots d-1$ and I have rescaled $\phi\rightarrow \phi \sqrt{T}$ and I have indicated the original dimension for the coupling $g_B$ by a subscript. Note that the mass term in (\ref{dimred}) need not coincide with $m_B^2$ of the original theory because additional contributions are typically generated when integrating out the ``static'' modes (cf. the discussion in Ref.~\cite{Laine:2016hma}). In this form $S_{\rm red}$ corresponds to (\ref{action}) at zero temperature in one dimension lower, and one can immediately relate the coupling constants
\be
\label{cmatching}
g_{B,(d-1)}=T g_{B,(d)}\,.
\ee
In order to related the mass parameter, one matches observables in the original d-dimensional theory at high temperature, and the dimensionally reduced theory. For instance, the let us consider the pole mass as an observable, which is determined by the solution to the saddle point condition. Matching the symmetric saddle point condition at high temperature  (\ref{d2highT}) in $d=2$ to the saddle point condition at zero temperature (\ref{d1res}) in $d=1$ with mass $m_{B,(1)}$ and coupling $g_{B,(1)}$ one finds
\be
\label{d1matching}
g_{B,(1)}=T g_{B,(2)}\,,\quad m_{B,(1)}^2=-\frac{3 g_{B,(2)}}{\pi}\ln \frac{\Lambda_{\overline{\rm MS},(2)}^2 e^{2 \gamma_E}}{(4 \pi T)^2}\,,
\ee
where I stress that $T$ here denotes the temperature in the $d=2$ dimensional theory, whereas the effective $d=1$ description is at zero temperature. Also, it is worth pointing out that the $d=2$ renormalization scale $\Lambda_{\overline{\rm MS},(2)}$ in this matching fulfills
\be
\label{lamd1}
\Lambda_{\overline{\rm MS},(2)}=4\pi T e^{-\gamma_E-\frac{\pi m_{B,(1)}^2 }{6 g_{B,(2)}}}\,,
\ee
e.g. is temperature-dependent for fixed $m_{B,(1)}^2$.

A completely analogous construction can be performed for the broken phase pole mass $\tilde M$, where one finds exactly the same matching conditions (\ref{d1matching}) as for the symmetric saddle.

With the parameters for the effective $d=1$ action fixed, one can directly reuse the results found in the previous section. One finds that for quantum mechanics (d=1) at zero temperature, the system is in the broken phase as long as
\be
\frac{m_{B,(1)}^2}{g_{B,(1)^{\frac{2}{3}}}}=-\frac{3 g_{B,(2)}^{\frac{1}{3}}}{\pi T^{\frac{2}{3}}}\ln \frac{\Lambda_{\overline{\rm MS},(2)}^2 e^{2 \gamma_E}}{(4 \pi T)^2}\lesssim 6.66\,,
\ee
which is always true in the high temperature limit. From this requirement, one can immediately obtain an estimate for the critical temperature $T_c$ in the $d=2$ theory as
\be
\label{d2hightap}
\frac{g_{B,(2)}}{\Lambda_{\overline{\rm MS},(2)}^2}\simeq \frac{6.66^3 \pi^3 T_c^2}{27 \ln\frac{(4 \pi T_c)^2}{e^{2\gamma_E}\Lambda_{\overline{\rm MS},(2)}^2}}\,,
\ee
which is also shown in Fig.\ref{fig:2d}.

\begin{figure}
  \includegraphics[width=.7\linewidth]{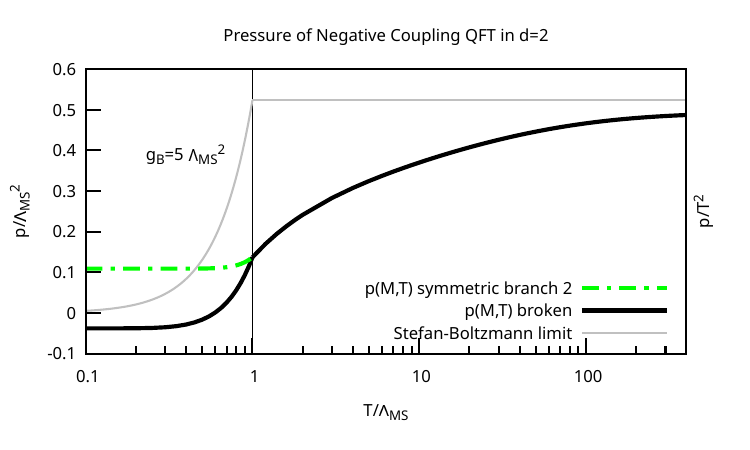}
  \caption{\label{fig:2dpressure} Pressure for the dominant phase as a function of temperature for $g_B=5 \Lambda_{\overline{\rm MS}}^2$. Left part of the plot shows $\frac{p(M,T)}{\Lambda_{\overline{\rm MS}}^2}$ whereas right part shows $\frac{p(M,T)}{T^2}$ to indicate approach to the Stefan-Boltzmann limit $p_{SB}(T)=\frac{\pi T^2}{6}$ (grey line).}
\end{figure}

The result at high temperature is also interesting because the solution to the symmetric phase saddle (\ref{d2highT}) becomes
\be
M=(-3 g_{B,(2)} T)^{\frac{1}{3}}\,,
\ee
which is not real-valued for the principal branch of the root. This leads to the well-documented case of complex-valued pressure functions at high temperature \cite{Ai:2022csx,Romatschke:2022jqg,Lawrence:2023woz,Kamata:2023opn,Romatschke:2024cld}. In the previous article \cite{Romatschke:2024cld}, I argued that the non-principal branch of the root function leads to a well-behaved high-temperature limit. In this work, the existence of a real-valued pole mass and pressure from the principal root of the broken saddle expansion, specifically
\be
\tilde M=(6 g_{B,(2)} T)^{\frac{1}{3}}\,,
\ee
make appeals to higher Riemann sheets unnecessary, and in my opinion provide a better resolution of the problems discussed in Refs.~\cite{Ai:2022csx,Romatschke:2022jqg,Lawrence:2023woz,Kamata:2023opn,Romatschke:2024cld}.

A representative plot for the pressure for the thermodynamically dominant phase as a function of temperature is shown in Fig.~\ref{fig:2dpressure} for $g_B=5\Lambda{\overline{\rm MS}}$. As can be seen from this figure, the symmetric branch is preferred at low temperatures, but there is a phase transition at $\frac{T_c}{\Lambda_{\overline{\rm MS}}}\simeq 1.$ For $T>T_c$, the broken saddle expansion provides a real-valued pressure that approaches the Stefan-Boltzmann limit in the high temperature limit.

\section{d=3: Negative Coupling Field Theory}

For the case of $d=3$, the symmetric phase and broken phase results lead to
\ba
\label{d3res}
p(M,T)&=&\frac{M^3}{12\pi}-\frac{(M^2-m_B^2)^2}{48g_B}+\frac{T^3}{2\pi}\left({\rm Li}_3\left(e^{-\beta M}\right)+\beta M {\rm Li}_2\left(e^{-\beta M}\right)\right)\,,\\
M^2&=&m_B^2+\frac{3 g_B M}{\pi}+\frac{6 g_B T}{\pi}\ln\left(1-e^{-\beta M}\right)\,,\nonumber\\
  \tilde p(\tilde M,T)&=&\frac{\tilde M^3}{12\pi}
  +\frac{\tilde M^4}{96g_B}+\frac{\tilde M^2 m_B^2}{24g_B}-\frac{m_B^4}{48g_B}+\frac{T^3}{2\pi}\left({\rm Li}_3\left(e^{-\beta \tilde M}\right)+\beta \tilde M {\rm Li}_2\left(e^{-\beta \tilde M}\right)\right)\,,\\
-\frac{\tilde M^2}{2}&=&m_B^2+\frac{3 g_B \tilde M}{\pi}+\frac{6 g_B T}{\pi}\ln\left(1-e^{-\beta \tilde M}\right)\,,\nonumber
\ea
where I note that because of the absence of logarithmic divergencies no non-trivial renormalization is necessary, and I used
\be
\int \frac{d^2k}{(2\pi)^2}\ln\left(1-e^{-\beta \sqrt{k^2+M^2}}\right)=-\frac{T^3}{(2\pi)} {\rm Li}_3\left(e^{-\beta M}\right)-\frac{M T^2}{(2\pi)} {\rm Li}_2\left(e^{-\beta M}\right)\,,
  \ee
  in terms of the polylogarithm functions ${\rm Li}_n$.

  In the zero temperature limit $T=0$, the saddle point equations admit solutions
  \be
  M=\frac{3 g_B}{2\pi}\left(1\pm \sqrt{1+\frac{4 \pi^2 m_B^2}{9 g_B^2}}\right)\,,\quad
  \tilde M=-\frac{3 g_B}{\pi}\left(1\pm \sqrt{1-\frac{2m_B^2\pi^2}{9 g_B^2}}\right)\,,\quad
  \ee
  where $M\in \mathbb{R}^+$ for at least one branch for $m_B^2\geq -\frac{9 g_B^2}{4\pi^2}$ and $\tilde M\in \mathbb{R}^+$ for the minus branch and $m_B^2<0$. One finds that $\Delta p(M,\tilde M,0)=0$ for a critical value of
  \be
  \label{3dcrit}
  \frac{m_B^2}{g_B^2}\simeq -0.21\,,\quad \frac{g_B^2}{M^2}\simeq 2.68\,,
  \ee
  which may lend itself to comparison to lattice-based approaches.

\begin{figure}
  \includegraphics[width=.48\linewidth]{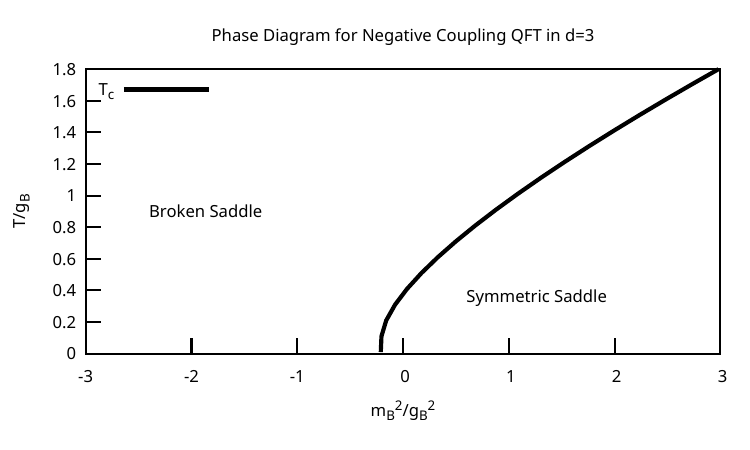}
 \hfill
 \includegraphics[width=.48\linewidth]{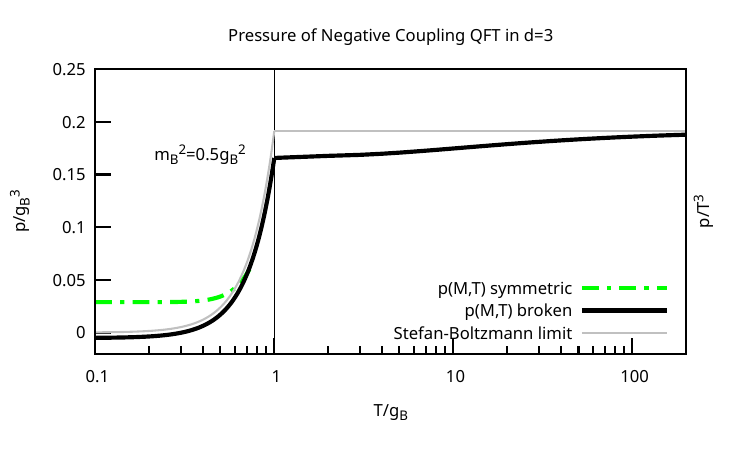}
  \caption{\label{fig:3d}
   Left: Phase diagram at finite temperature, indicating which saddle is thermodynamically preferred. Right: pressure as a function of temperature for $m_B^2=0.5 g_B^2$. Left part of the plot shows $\frac{p(M,T)}{g_B^3}$ whereas right part shows $\frac{p(M,T)}{T^3}$ to indicate approach to the Stefan-Boltzmann limit $p_{SB}(T)=\frac{\zeta(3) T^3}{2\pi}$ (grey line). For $m_B^2=0.5 g_B^2$, the broken saddle pressure exceeds the symmetric saddle branch above $T_c\simeq 0.71 g_B$. See text for details.}
\end{figure}
  
  \subsection{Nonzero temperature and high temperature limit}
  
  One can track the transition between the symmetric and broken saddle by following the line defined by $\Delta p(M,\tilde M,T)=0$, which is shown in Fig.~\ref{fig:3d}. In the high temperature limit, one can use the expansions (\ref{htep}) or more straightforwardly directly expand (\ref{d3res}) to find
  \ba
  \label{d3highT}
 p(M,T)=\frac{\zeta(3) T^3}{2\pi}-\frac{M^2 T}{8\pi}\left(1-2\ln(\beta M)\right)\,,\quad
M^2=\frac{6 g_B T}{\pi}\ln\left(\beta M\right)\,,\nonumber\\
  \tilde p(\tilde M,T)=\frac{\zeta(3) T^3}{2\pi}-\frac{\tilde M^2 T}{8\pi}\left(1-2\ln(\beta \tilde M)\right)\,,\quad
-\frac{\tilde M^2}{2}=\frac{6 g_B T}{\pi}\ln\left(\beta \tilde M\right)\,,\nonumber
\ea
which have solutions
\be
M^2=-\frac{3 g_{B} T}{\pi} W_{0,-1}\left(-\frac{\pi T}{3 g_B}\right)\,,\quad
\tilde M^2=\frac{6 g_{B} T}{\pi} W_{0}\left(\frac{\pi T}{6 g_B}\right)\,,
\ee
for the pole masses of the symmetric and broken saddle expansions. Using the matching condition (\ref{cmatching}) and comparing the pole masses to the zero-temperature expressions (\ref{d2t0gap},\ref{d2t0tildegap}) one finds
\be
g_{B,(2)}=T g_{B,(3)}\,,\quad \Lambda_{\overline{\rm MS},(2)}=T\,.
\ee
Note that in particular the renormalization scale $\Lambda_{\overline{\rm MS}}$ for the two-dimensionally reduced theory takes the value of the temperature of the three-dimensional theory. As was found before, the matching conditions are independent of matching the symmetric or broken saddle expansion the dimensionally reduced theory.

However, in the high temperature limit the symmetric saddle solution $M$ as well as the corresponding pressure becomes complex-valued, see the discussion for $d=2$ around Eq.~(\ref{2dcomplex}). The solution is indicated in Fig.~\ref{fig:3d}: rather than tracking the symmetric saddle, at high temperatures a real-valued solution is provided by the broken saddle expansion.

A representative example of the pressure as a function of temperature is shown in Fig.~\ref{fig:3d} for the case of $m_B^2=0.5 g_B^2$. For low temperatures, the dominant phase is given by the symmetric saddle expansion until a critical temperature of $T_c\simeq 0.71 g_B$, above which the broken phase saddle provides higher pressure. At high temperatures, the broken phase saddle pressure approaches the Stefan-Boltzmann pressure asymptotically from below.

\section{d=4: Non-trivial Interacting Scalar Field Theory}

Let me now consider the case $d=4$. Before presenting the results, let me preface the discussion by pointing out that there is a mathematical proof that scalar field theory in four dimensions is trivial in the continuum, ruling out interacting field theory in four dimensions \cite{Aizenman:2019yuo}. The mathematical proof is very specific, and in particular is only valid for Euclidean actions that are within the Griffiths-Simon class. Notably, the case of negative coupling $\lambda=-g$ with $g\in \mathbb{R}^+$ is \textbf{not} in the Griffiths-Simon class, providing a loophole for which the mathematical proof in Ref.~\cite{Aizenman:2019yuo} does not apply \cite{Romatschke:2023ogd}, which has been pointed out before in Ref.~\cite{1983NuPhB.225..261A}. Similarly, lattice-based arguments against an interacting continuum theory from the 1980s \cite{Frohlich:1982tw,Wilson:1983xri} suffer from the same loophole in that positive coupling was implicitly assumed as a sine-qua-non condition. Lattice studies for the negative coupling theory using contour deformations and brute force numerical integration have only been performed recently \cite{Romatschke:2023fax}, but only for tiny lattices, preventing the study of the continuum theory using this method.

Since no known result in the literature prohibits an interacting continuum scalar field theory in $d=4-2\varepsilon$ with $\varepsilon\rightarrow 0$  and negative coupling \cite{1983NuPhB.225..261A}, I now proceed to evaluate precisely this case using the saddle point expansions from above. I find
\ba
\label{d4res}
p(M,T)&=&\frac{M^4}{64\pi^2}\left(\frac{1}{\varepsilon}+\ln\frac{\bar\mu^2 e^{\frac{3}{2}}}{M^2}\right)-\frac{(M^2-m_B^2)^2}{48g_B}+\frac{M^2 T^2}{2\pi^2}\sum_{n=1}^\infty \frac{K_{2}(n \beta M)}{n^2}\,,\\
M^2&=&m_B^2+\frac{3M^2 g_B}{4\pi^2}\left(\frac{1}{\varepsilon}+\ln\frac{\bar\mu^2 e^{1}}{M^2}\right)-\frac{6 g_B M T}{\pi^2}\sum_{n=1}^\infty \frac{K_1(\beta M n)}{n}\,,\nonumber\\
\tilde p(\tilde M,T)&=&\frac{\tilde M^4}{64\pi^2}\left(\frac{1}{\varepsilon}+\ln\frac{\bar\mu^2 e^{\frac{3}{2}}}{\tilde M^2}\right)+\frac{\tilde M^4}{96g_B}+\frac{\tilde M^2 m_B^2}{24g_B}-\frac{m_B^4}{48g_B}+\frac{\tilde M^2 T^2}{2\pi^2}\sum_{n=1}^\infty \frac{K_{2}(n \beta \tilde M)}{n^2}\,,\\
-\frac{\tilde M^2}{2}&=&m_B^2+\frac{3\tilde M^2 g_B}{4\pi^2}\left(\frac{1}{\varepsilon}+\ln\frac{\bar\mu^2 e^{1}}{\tilde M^2}\right)-\frac{6 g_B \tilde M T}{\pi^2}\sum_{n=1}^\infty \frac{K_1(\beta \tilde M n)}{n}\,,\nonumber
\ea
where I used
\be
\int \frac{dk}{2\pi^2} k^2 \ln\left(1-e^{-\beta \sqrt{k^2+M^2}}\right)=-\frac{M^2 T}{2\pi^2}\sum_{n=1}^\infty \frac{K_{2}(n \beta M)}{n^2}\,.
\ee

The symmetric saddle can be non-perturbatively renormalized as
\be
\label{ren4d1}
\frac{1}{g_B}=\frac{1}{g_R(\bar\mu)}+\frac{3}{4\pi^2\varepsilon},\quad \frac{1}{g_R(\bar\mu)}=\frac{3}{4\pi^2}\ln \frac{\bar\mu^2}{\Lambda_{\overline{\rm MS}}^2}\,,
\ee
whereas the broken saddle can be non-perturbatively renormalized as
\be
\label{ren4d2}
\frac{1}{g_B}=\frac{1}{g_R(\bar\mu)}-\frac{3}{2\pi^2\varepsilon},\quad \frac{1}{\tilde g_R(\bar\mu)}=\frac{3}{2\pi^2}\ln \frac{\tilde \Lambda_{\overline{\rm MS}}^2}{\bar\mu^2}\,,
\ee
see the discussions in Ref.~\cite{Romatschke:2026tam}. Note that unlike in lower dimensions, symmetric and broken saddle cannot be simultaneously renormalized using the same renormalization condition, which introduces the need for two scales: $\Lambda_{\overline{\rm MS}}^2,\tilde \Lambda_{\overline{\rm MS}}^2$. However, after renormalization, (\ref{d4res}) becomes 
\ba
\label{d4res2}
p(M,T)&=&\frac{M^4}{64\pi^2}\ln\frac{\Lambda_{\overline{\rm MS}}^2 e^{\frac{3}{2}}}{M^2}+\frac{M^2 m_B^2}{24 g_B}+\frac{M^2 T^2}{2\pi^2}\sum_{n=1}^\infty \frac{K_{2}(n \beta M)}{n^2}\,,\\
0&=&\frac{m_B^2}{g_B} +\frac{3M^2}{4\pi^2}\ln\frac{\Lambda_{\overline{\rm MS}}^2 e^{1}}{M^2}-\frac{6 M T}{\pi^2}\sum_{n=1}^\infty \frac{K_1(\beta M n)}{n}\,,\nonumber\\
\tilde p(\tilde M,T)&=&\frac{\tilde M^4}{64\pi^2}\ln\frac{\tilde \Lambda_{\overline{\rm MS}}^2 e^{\frac{3}{2}}}{\tilde M^2}+\frac{\tilde M^2m_B^2 }{24g_B}+\frac{\tilde M^2 T^2}{2\pi^2}\sum_{n=1}^\infty \frac{K_{2}(n \beta \tilde M)}{n^2}\,,\\
0&=&\frac{m_B^2}{g_B}+\frac{3\tilde M^2}{4\pi^2}\ln\frac{\tilde \Lambda_{\overline{\rm MS}}^2 e^{1}}{\tilde M^2}-\frac{6 \tilde M T}{\pi^2}\sum_{n=1}^\infty \frac{K_1(\beta \tilde M n)}{n}\,,\nonumber
\ea
implying that the free energy and pole masses in the symmetric and broken saddle expansion would be exactly identical if $\Lambda_{\overline{\rm MS}}=\tilde \Lambda_{\overline{\rm MS}}$.

However, for $\Lambda_{\overline{\rm MS}}=\tilde \Lambda_{\overline{\rm MS}}$ one has different values of $g_B$ in the broken and symmetric phase, contradicting the derivation from the same action (\ref{action}).

A partial resolution is indicated by considering the high temperature limit of the theory provided by (\ref{htep}) which for the case at hand become after renormalization
\ba
\label{d4res3}
p(M,T)&=&\frac{\pi^2 T^4}{90}-\frac{M^2T^2}{24}+\frac{M^3 T}{12\pi}+\frac{M^4}{64\pi^2}\ln\frac{\Lambda_{\overline{\rm MS}}^2e^{2\gamma_E}}{(4 \pi T)^2}\,,\quad 0=\frac{3M^2}{4\pi^2}\ln\frac{\Lambda_{\overline{\rm MS}}^2e^{2\gamma_E}}{(4\pi T)^2}-T^2+\frac{3 M T}{\pi}\,,\\
\tilde p(\tilde M,T)&=&\frac{\pi^2 T^4}{90}-\frac{\tilde M^2T^2}{24}+\frac{\tilde M^3 T}{12\pi}+\frac{\tilde M^4}{64\pi^2}\ln\frac{\tilde \Lambda_{\overline{\rm MS}}^2e^{2\gamma_E}}{(4 \pi T)^2}\,,\quad 0=\frac{3\tilde M^2}{4\pi^2}\ln\frac{\tilde \Lambda_{\overline{\rm MS}}^2e^{2\gamma_E}}{(4\pi T)^2}-T^2+\frac{3 \tilde M T}{\pi}\,,
\ea

Comparing the saddle-point conditions for $M,\tilde M$ with the effective dimensionally reduced theory (\ref{d3res}) leads to
\ba
g_{B,(3)}=-\frac{4 \pi^2 T}{3\ln\frac{\Lambda_{\overline{\rm MS}}^2e^{2\gamma_E}}{(4\pi T)^2}}\,,\quad m_{B,(3)}^2=-T g_{B,(3)}\,,\\
\tilde g_{B,(3)}=\frac{2 \pi^2 T}{3\ln\frac{\tilde \Lambda_{\overline{\rm MS}}^2e^{2\gamma_E}}{(4\pi T)^2}}\,,\quad m_{B,(3)}^2=-T \tilde g_{B,(3)}\,.
\ea
The effective three dimensional theories for the symmetric and broken saddle expansion become identical if $g_{B,(3)}=\tilde g_{B,(3)}$ or
\be
\label{m4d}
\tilde \Lambda_{\overline{\rm MS}}^2=\frac{(4 \pi T)^3}{\Lambda_{\overline{\rm MS}} e^{3\gamma_E}}\,.
\ee
Note the similarity to (\ref{lamd1}).

With the effective parameters for the three-dimensional theory fixed, one can use (\ref{3dcrit}) to determine the approximate transition temperature $T_c$ from broken to symmetric saddle as
\be
\frac{m_{B,(3)}^2}{g_{B,(3)}^2}=\frac{3\ln\frac{\Lambda_{\overline{\rm MS}}^2e^{2\gamma_E}}{(4\pi T_c)^2}}{4 \pi^2 }\simeq -0.21\,,
\ee
implying $T_c\simeq 0.564 \Lambda_{\overline{\rm MS}}$. The saddle-point conditions (\ref{d4res3}) in the high temperature limit have the solutions
\be
M= -\frac{2\pi T}{\ln\frac{\Lambda_{\overline{\rm MS}}^2e^{2\gamma_E}}{(4\pi T)^2}}\left(1\pm \sqrt{1+\frac{1}{3}\ln\frac{\Lambda_{\overline{\rm MS}}^2e^{2\gamma_E}}{(4\pi T)^2}}\right)\,,\quad \tilde M= \frac{4\pi T}{\ln\frac{\Lambda_{\overline{\rm MS}}^2e^{2\gamma_E}}{(4\pi T)^2}}\left(1\pm \sqrt{1-\frac{1}{6}\ln\frac{\Lambda_{\overline{\rm MS}}^2e^{2\gamma_E}}{(4\pi T)^2}}\right)\,,
\ee
where the only solution $\in \mathbb{R}^+$ is
\be
\tilde M=\frac{4\pi T}{\ln\frac{\Lambda_{\overline{\rm MS}}^2e^{2\gamma_E}}{(4\pi T)^2}}\left(1- \sqrt{1-\frac{1}{6}\ln\frac{\Lambda_{\overline{\rm MS}}^2e^{2\gamma_E}}{(4\pi T)^2}}\right)\,.
\label{sol4d}
\ee
In particular, the solution for the symmetric saddle mass $M$ is complex-valued, and the resulting pressure is complex-valued as well, which has created much discussion in the recent literature \cite{Ai:2022csx,Romatschke:2022jqg,Kamata:2023opn,Lawrence:2023woz,Kamata:2024tyb,Romatschke:2024cld}.

The existence of the real-valued broken-saddle solution (\ref{sol4d}) implies a real-valued pressure for the high-temperature limit of negative coupling field theory, thereby resolving the issue of complex-valued pressure for the four-dimensional scalar field theory.

At zero temperature, the solution to the symmetric phase saddle-point condition (\ref{d4res2}) is
\be
M^2=\frac{4 \pi^2 m_B^2}{3 g_B W_0\left(\frac{4 m_B^2 \pi^2}{3e^1g_B \Lambda_{\overline{\rm MS}}^2}\right)}\,,
\ee
which implies $M\in \mathbb{R}^+$ for $\frac{m_B^2}{g_B \Lambda_{\overline{\rm MS}}^2}\geq -\frac{3}{4\pi^2}$. For the broken phase, using the renormalization prescription (\ref{ren4d2}) implies
\be
\tilde M=0\,,\quad \forall \frac{m_B^2}{g_B \Lambda_{\overline{\rm MS}}^2}={\rm finite}
\ee
and also $\tilde p(\tilde M,T=0)=0$. This suggests that at low values of $\frac{m_B^2}{g_B \Lambda_{\overline{\rm MS}}^2}$, the broken phase is dominant, whereas for sufficiently high $\frac{m_B^2}{g_B}$ the symmetric phase is dominant. Increasing $m_B^2$ from $\frac{m_B^2}{g_B \Lambda_{\overline{\rm MS}}^2}=-\frac{3}{4\pi^2}$, one finds indeed that $\Delta p(M,\tilde M,T=0)=0$ at
\be
\frac{m_B^2}{g_B\Lambda_{\overline{\rm MS}}^2}\simeq -0.06264\,,
\ee
suggesting a phase transition from the broken to symmetric saddle for this value of parameters.

\begin{figure}
  \includegraphics[width=.7\linewidth]{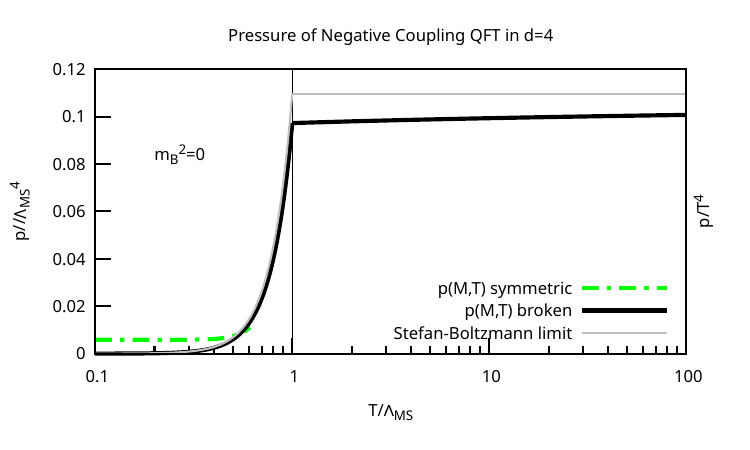}
  \caption{\label{fig:4dpressure} Pressure as a function of temperature for $m_B^2=0$. Left part of the plot shows $\frac{p(M,T)}{\Lambda_{\overline{\rm MS}}^4}$ whereas right part shows $\frac{p(M,T)}{T^4}$ to indicate approach to the Stefan-Boltzmann limit $p_{SB}(T)=\frac{\pi^2 T^4}{90}$ (grey line). See text for details.}
\end{figure}

At finite temperature and $m_B^2=0$, using (\ref{m4d}) directly in (\ref{d4res}) leads to two solutions for $\tilde M$, the smaller one of which reproduces (\ref{sol4d}) in the high temperature limit. A plot for the pressure in the symmetric phase and broken phase tracking the smaller solution for $\tilde M$ is shown in Fig.~\ref{fig:4dpressure}.

As can be seen in Fig.~\ref{fig:4dpressure}, the pressure is well-defined for all temperatures and approaches the Stefan Boltzmann limit asymptotically from below. One finds $\Delta p(M,\tilde M,T_c)=0$ for $T_c\simeq 0.543 \Lambda_{\overline{\rm MS}}$, which is well below the temperature $T\simeq 0.616 \Lambda_{\overline{\rm MS}}$ where the symmetric pole mass $M$ becomes complex valued \cite{Romatschke:2022jqg}.

\section{Summary and Conclusions}

In this work I have considered scalar field theory with negative quartic coupling using saddle point expansions in various dimensions. It was found that the analytic saddle point expansions predict an interesting phase diagram structure for the continuum theory, which persists at finite temperature. For the case of quantum mechanics at negative coupling, which is alternatively accessible through its isospectral Hermitian equivalent \cite{Jones:2006qs}, the saddle point expansions were found to be quantitatively reliable, agreeing with the numerically diagonalized Hamiltonian on the level of 15 percent. In two dimensions, the saddle point expansions point to a phase transition at zero temperature that may be probed by numerical methods in the near future, such as brute-force numerical integration on the lattice \cite{Romatschke:2023fax}. At finite temperature, the issue of complex-valued pressure reported previously for negative coupling field theory \cite{Romatschke:2024cld} is resolved through a phase transition to the broken phase saddle, leading to a well-defined phase structure of the theory for all temperatures.

For three dimensions, the saddle point expansions likewise suggest a phase transition at zero temperature located near bare parameter values $\frac{m_B^2}{g_B}\simeq -0.21$, and similarly a well-defined phase structure for the theory at finite temperature.

The case of four dimensions is considerably more complex, but the dimensionally reduced theory at high temperatures points to a well-defined real-valued pressure function described by the broken phase saddle, while at zero temperature and positive $m_B^2$ the dominant thermodynamic configuration is described by the symmetric phase saddle. Nevertheless, the fact that negative coupling exploits a known loophole in quantum triviality proofs \cite{Aizenman:2019yuo,1983NuPhB.225..261A,Romatschke:2023ogd} and suggests asymptotic freedom for scalars \cite{Symanzik:1973hx,Romatschke:2023sce} makes further study of this case particularly interesting for the future.

\bibliography{enormous}

\begin{thebibliography}{54}%
\makeatletter
\providecommand \@ifxundefined [1]{%
 \@ifx{#1\undefined}
}%
\providecommand \@ifnum [1]{%
 \ifnum #1\expandafter \@firstoftwo
 \else \expandafter \@secondoftwo
 \fi
}%
\providecommand \@ifx [1]{%
 \ifx #1\expandafter \@firstoftwo
 \else \expandafter \@secondoftwo
 \fi
}%
\providecommand \natexlab [1]{#1}%
\providecommand \enquote  [1]{``#1''}%
\providecommand \bibnamefont  [1]{#1}%
\providecommand \bibfnamefont [1]{#1}%
\providecommand \citenamefont [1]{#1}%
\providecommand \href@noop [0]{\@secondoftwo}%
\providecommand \href [0]{\begingroup \@sanitize@url \@href}%
\providecommand \@href[1]{\@@startlink{#1}\@@href}%
\providecommand \@@href[1]{\endgroup#1\@@endlink}%
\providecommand \@sanitize@url [0]{\catcode `\\12\catcode `\$12\catcode
  `\&12\catcode `\#12\catcode `\^12\catcode `\_12\catcode `\%12\relax}%
\providecommand \@@startlink[1]{}%
\providecommand \@@endlink[0]{}%
\providecommand \url  [0]{\begingroup\@sanitize@url \@url }%
\providecommand \@url [1]{\endgroup\@href {#1}{\urlprefix }}%
\providecommand \urlprefix  [0]{URL }%
\providecommand \Eprint [0]{\href }%
\providecommand \doibase [0]{https://doi.org/}%
\providecommand \selectlanguage [0]{\@gobble}%
\providecommand \bibinfo  [0]{\@secondoftwo}%
\providecommand \bibfield  [0]{\@secondoftwo}%
\providecommand \translation [1]{[#1]}%
\providecommand \BibitemOpen [0]{}%
\providecommand \bibitemStop [0]{}%
\providecommand \bibitemNoStop [0]{.\EOS\space}%
\providecommand \EOS [0]{\spacefactor3000\relax}%
\providecommand \BibitemShut  [1]{\csname bibitem#1\endcsname}%
\let\auto@bib@innerbib\@empty
\bibitem [{\citenamefont {Symanzik}(1973)}]{Symanzik:1973hx}%
  \BibitemOpen
  \bibfield  {author} {\bibinfo {author} {\bibfnamefont {K.}~\bibnamefont
  {Symanzik}},\ }\bibfield  {title} {\bibinfo {title} {{A field theory with
  computable large-momenta behavior}},\ }\href
  {https://doi.org/10.1007/BF02788323} {\bibfield  {journal} {\bibinfo
  {journal} {Lett. Nuovo Cim.}\ }\textbf {\bibinfo {volume} {6S2}},\ \bibinfo
  {pages} {77} (\bibinfo {year} {1973})}\BibitemShut {NoStop}%
\bibitem [{\citenamefont {Coleman}\ and\ \citenamefont
  {Gross}(1973)}]{Coleman:1973sx}%
  \BibitemOpen
  \bibfield  {author} {\bibinfo {author} {\bibfnamefont {S.~R.}\ \bibnamefont
  {Coleman}}\ and\ \bibinfo {author} {\bibfnamefont {D.~J.}\ \bibnamefont
  {Gross}},\ }\bibfield  {title} {\bibinfo {title} {{Price of asymptotic
  freedom}},\ }\href {https://doi.org/10.1103/PhysRevLett.31.851} {\bibfield
  {journal} {\bibinfo  {journal} {Phys. Rev. Lett.}\ }\textbf {\bibinfo
  {volume} {31}},\ \bibinfo {pages} {851} (\bibinfo {year} {1973})}\BibitemShut
  {NoStop}%
\bibitem [{\citenamefont {Bender}\ and\ \citenamefont
  {Boettcher}(1998)}]{Bender:1998ke}%
  \BibitemOpen
  \bibfield  {author} {\bibinfo {author} {\bibfnamefont {C.~M.}\ \bibnamefont
  {Bender}}\ and\ \bibinfo {author} {\bibfnamefont {S.}~\bibnamefont
  {Boettcher}},\ }\bibfield  {title} {\bibinfo {title} {{Real spectra in
  nonHermitian Hamiltonians having PT symmetry}},\ }\href
  {https://doi.org/10.1103/PhysRevLett.80.5243} {\bibfield  {journal} {\bibinfo
   {journal} {Phys. Rev. Lett.}\ }\textbf {\bibinfo {volume} {80}},\ \bibinfo
  {pages} {5243} (\bibinfo {year} {1998})},\ \Eprint
  {https://arxiv.org/abs/physics/9712001} {arXiv:physics/9712001} \BibitemShut
  {NoStop}%
\bibitem [{\citenamefont {Bender}\ \emph {et~al.}(1999)\citenamefont {Bender},
  \citenamefont {Boettcher},\ and\ \citenamefont {Meisinger}}]{bender1999}%
  \BibitemOpen
  \bibfield  {author} {\bibinfo {author} {\bibfnamefont {C.~M.}\ \bibnamefont
  {Bender}}, \bibinfo {author} {\bibfnamefont {S.}~\bibnamefont {Boettcher}},\
  and\ \bibinfo {author} {\bibfnamefont {P.~N.}\ \bibnamefont {Meisinger}},\
  }\bibfield  {title} {\bibinfo {title} {Pt-symmetric quantum mechanics},\
  }\href@noop {} {\bibfield  {journal} {\bibinfo  {journal} {Journal of
  Mathematical Physics}\ }\textbf {\bibinfo {volume} {40}},\ \bibinfo {pages}
  {2201} (\bibinfo {year} {1999})}\BibitemShut {NoStop}%
\bibitem [{\citenamefont {Dorey}\ \emph {et~al.}(2001)\citenamefont {Dorey},
  \citenamefont {Dunning},\ and\ \citenamefont {Tateo}}]{Dorey:2001uw}%
  \BibitemOpen
  \bibfield  {author} {\bibinfo {author} {\bibfnamefont {P.}~\bibnamefont
  {Dorey}}, \bibinfo {author} {\bibfnamefont {C.}~\bibnamefont {Dunning}},\
  and\ \bibinfo {author} {\bibfnamefont {R.}~\bibnamefont {Tateo}},\ }\bibfield
   {title} {\bibinfo {title} {{Spectral equivalences, Bethe Ansatz equations,
  and reality properties in PT-symmetric quantum mechanics}},\ }\href
  {https://doi.org/10.1088/0305-4470/34/28/305} {\bibfield  {journal} {\bibinfo
   {journal} {J. Phys. A}\ }\textbf {\bibinfo {volume} {34}},\ \bibinfo {pages}
  {5679} (\bibinfo {year} {2001})},\ \Eprint
  {https://arxiv.org/abs/hep-th/0103051} {arXiv:hep-th/0103051} \BibitemShut
  {NoStop}%
\bibitem [{\citenamefont {Bender}(2007)}]{Bender:2007nj}%
  \BibitemOpen
  \bibfield  {author} {\bibinfo {author} {\bibfnamefont {C.~M.}\ \bibnamefont
  {Bender}},\ }\bibfield  {title} {\bibinfo {title} {{Making sense of
  non-Hermitian Hamiltonians}},\ }\href
  {https://doi.org/10.1088/0034-4885/70/6/R03} {\bibfield  {journal} {\bibinfo
  {journal} {Rept. Prog. Phys.}\ }\textbf {\bibinfo {volume} {70}},\ \bibinfo
  {pages} {947} (\bibinfo {year} {2007})},\ \Eprint
  {https://arxiv.org/abs/hep-th/0703096} {arXiv:hep-th/0703096} \BibitemShut
  {NoStop}%
\bibitem [{\citenamefont {Dorey}\ \emph {et~al.}(2007)\citenamefont {Dorey},
  \citenamefont {Dunning},\ and\ \citenamefont {Tateo}}]{Dorey:2007zx}%
  \BibitemOpen
  \bibfield  {author} {\bibinfo {author} {\bibfnamefont {P.}~\bibnamefont
  {Dorey}}, \bibinfo {author} {\bibfnamefont {C.}~\bibnamefont {Dunning}},\
  and\ \bibinfo {author} {\bibfnamefont {R.}~\bibnamefont {Tateo}},\ }\bibfield
   {title} {\bibinfo {title} {{The ODE/IM Correspondence}},\ }\href
  {https://doi.org/10.1088/1751-8113/40/32/R01} {\bibfield  {journal} {\bibinfo
   {journal} {J. Phys. A}\ }\textbf {\bibinfo {volume} {40}},\ \bibinfo {pages}
  {R205} (\bibinfo {year} {2007})},\ \Eprint
  {https://arxiv.org/abs/hep-th/0703066} {arXiv:hep-th/0703066} \BibitemShut
  {NoStop}%
\bibitem [{\citenamefont {{El-Ganainy}}\ \emph {et~al.}(2018)\citenamefont
  {{El-Ganainy}}, \citenamefont {{Makris}}, \citenamefont {{Khajavikhan}},
  \citenamefont {{Musslimani}}, \citenamefont {{Rotter}},\ and\ \citenamefont
  {{Christodoulides}}}]{2018NatPh..14...11E}%
  \BibitemOpen
  \bibfield  {author} {\bibinfo {author} {\bibfnamefont {R.}~\bibnamefont
  {{El-Ganainy}}}, \bibinfo {author} {\bibfnamefont {K.~G.}\ \bibnamefont
  {{Makris}}}, \bibinfo {author} {\bibfnamefont {M.}~\bibnamefont
  {{Khajavikhan}}}, \bibinfo {author} {\bibfnamefont {Z.~H.}\ \bibnamefont
  {{Musslimani}}}, \bibinfo {author} {\bibfnamefont {S.}~\bibnamefont
  {{Rotter}}},\ and\ \bibinfo {author} {\bibfnamefont {D.~N.}\ \bibnamefont
  {{Christodoulides}}},\ }\bibfield  {title} {\bibinfo {title} {{Non-Hermitian
  physics and PT symmetry}},\ }\href {https://doi.org/10.1038/nphys4323}
  {\bibfield  {journal} {\bibinfo  {journal} {Nature Physics}\ }\textbf
  {\bibinfo {volume} {14}},\ \bibinfo {pages} {11} (\bibinfo {year}
  {2018})}\BibitemShut {NoStop}%
\bibitem [{\citenamefont {{R{\"u}ter}}\ \emph {et~al.}(2010)\citenamefont
  {{R{\"u}ter}}, \citenamefont {{Makris}}, \citenamefont {{El-Ganainy}},
  \citenamefont {{Christodoulides}}, \citenamefont {{Segev}},\ and\
  \citenamefont {{Kip}}}]{2010NatPh...6..192R}%
  \BibitemOpen
  \bibfield  {author} {\bibinfo {author} {\bibfnamefont {C.~E.}\ \bibnamefont
  {{R{\"u}ter}}}, \bibinfo {author} {\bibfnamefont {K.~G.}\ \bibnamefont
  {{Makris}}}, \bibinfo {author} {\bibfnamefont {R.}~\bibnamefont
  {{El-Ganainy}}}, \bibinfo {author} {\bibfnamefont {D.~N.}\ \bibnamefont
  {{Christodoulides}}}, \bibinfo {author} {\bibfnamefont {M.}~\bibnamefont
  {{Segev}}},\ and\ \bibinfo {author} {\bibfnamefont {D.}~\bibnamefont
  {{Kip}}},\ }\bibfield  {title} {\bibinfo {title} {{Observation of parity-time
  symmetry in optics}},\ }\href {https://doi.org/10.1038/nphys1515} {\bibfield
  {journal} {\bibinfo  {journal} {Nature Physics}\ }\textbf {\bibinfo {volume}
  {6}},\ \bibinfo {pages} {192} (\bibinfo {year} {2010})}\BibitemShut {NoStop}%
\bibitem [{\citenamefont {{Droulias}}\ \emph {et~al.}(2019)\citenamefont
  {{Droulias}}, \citenamefont {{Katsantonis}}, \citenamefont {{Kafesaki}},
  \citenamefont {{Soukoulis}},\ and\ \citenamefont
  {{Economou}}}]{2019PhRvL.122u3201D}%
  \BibitemOpen
  \bibfield  {author} {\bibinfo {author} {\bibfnamefont {S.}~\bibnamefont
  {{Droulias}}}, \bibinfo {author} {\bibfnamefont {I.}~\bibnamefont
  {{Katsantonis}}}, \bibinfo {author} {\bibfnamefont {M.}~\bibnamefont
  {{Kafesaki}}}, \bibinfo {author} {\bibfnamefont {C.~M.}\ \bibnamefont
  {{Soukoulis}}},\ and\ \bibinfo {author} {\bibfnamefont {E.~N.}\ \bibnamefont
  {{Economou}}},\ }\bibfield  {title} {\bibinfo {title} {{Chiral Metamaterials
  with P T Symmetry and Beyond}},\ }\href
  {https://doi.org/10.1103/PhysRevLett.122.213201} {\bibfield  {journal}
  {\bibinfo  {journal} {\prl}\ }\textbf {\bibinfo {volume} {122}},\ \bibinfo
  {eid} {213201} (\bibinfo {year} {2019})},\ \Eprint
  {https://arxiv.org/abs/1811.05344} {arXiv:1811.05344 [physics.optics]}
  \BibitemShut {NoStop}%
\bibitem [{\citenamefont {Fisher}(1978)}]{PhysRevLett.40.1610}%
  \BibitemOpen
  \bibfield  {author} {\bibinfo {author} {\bibfnamefont {M.~E.}\ \bibnamefont
  {Fisher}},\ }\bibfield  {title} {\bibinfo {title} {Yang-lee edge singularity
  and ${\ensuremath{\phi}}^{3}$ field theory},\ }\href
  {https://doi.org/10.1103/PhysRevLett.40.1610} {\bibfield  {journal} {\bibinfo
   {journal} {Phys. Rev. Lett.}\ }\textbf {\bibinfo {volume} {40}},\ \bibinfo
  {pages} {1610} (\bibinfo {year} {1978})}\BibitemShut {NoStop}%
\bibitem [{\citenamefont {Cardy}(1985)}]{PhysRevLett.54.1354}%
  \BibitemOpen
  \bibfield  {author} {\bibinfo {author} {\bibfnamefont {J.~L.}\ \bibnamefont
  {Cardy}},\ }\bibfield  {title} {\bibinfo {title} {Conformal invariance and
  the yang-lee edge singularity in two dimensions},\ }\href
  {https://doi.org/10.1103/PhysRevLett.54.1354} {\bibfield  {journal} {\bibinfo
   {journal} {Phys. Rev. Lett.}\ }\textbf {\bibinfo {volume} {54}},\ \bibinfo
  {pages} {1354} (\bibinfo {year} {1985})}\BibitemShut {NoStop}%
\bibitem [{\citenamefont {Li}\ and\ \citenamefont {Wang}(2025)}]{Li:2024xms}%
  \BibitemOpen
  \bibfield  {author} {\bibinfo {author} {\bibfnamefont {Y.-D.}\ \bibnamefont
  {Li}}\ and\ \bibinfo {author} {\bibfnamefont {Q.}~\bibnamefont {Wang}},\
  }\bibfield  {title} {\bibinfo {title} {{Isospectral local Hermitian theory
  for the PT-symmetric i{\ensuremath{\phi}}3 quantum field theory}},\ }\href
  {https://doi.org/10.1103/PhysRevD.111.025016} {\bibfield  {journal} {\bibinfo
   {journal} {Phys. Rev. D}\ }\textbf {\bibinfo {volume} {111}},\ \bibinfo
  {pages} {025016} (\bibinfo {year} {2025})},\ \Eprint
  {https://arxiv.org/abs/2412.10732} {arXiv:2412.10732 [hep-th]} \BibitemShut
  {NoStop}%
\bibitem [{\citenamefont {Arguello~Cruz}\ \emph {et~al.}(2026)\citenamefont
  {Arguello~Cruz}, \citenamefont {Klebanov}, \citenamefont {Tarnopolsky},\ and\
  \citenamefont {Xin}}]{ArguelloCruz:2025zuq}%
  \BibitemOpen
  \bibfield  {author} {\bibinfo {author} {\bibfnamefont {E.}~\bibnamefont
  {Arguello~Cruz}}, \bibinfo {author} {\bibfnamefont {I.~R.}\ \bibnamefont
  {Klebanov}}, \bibinfo {author} {\bibfnamefont {G.}~\bibnamefont
  {Tarnopolsky}},\ and\ \bibinfo {author} {\bibfnamefont {Y.}~\bibnamefont
  {Xin}},\ }\bibfield  {title} {\bibinfo {title} {{Yang-Lee Quantum Criticality
  in Various Dimensions}},\ }\href {https://doi.org/10.1103/w4qg-2xwn}
  {\bibfield  {journal} {\bibinfo  {journal} {Phys. Rev. X}\ }\textbf {\bibinfo
  {volume} {16}},\ \bibinfo {pages} {011022} (\bibinfo {year} {2026})},\
  \Eprint {https://arxiv.org/abs/2505.06369} {arXiv:2505.06369 [hep-th]}
  \BibitemShut {NoStop}%
\bibitem [{\citenamefont {Bender}\ \emph {et~al.}(2018)\citenamefont {Bender},
  \citenamefont {Hassanpour}, \citenamefont {Klevansky},\ and\ \citenamefont
  {Sarkar}}]{Bender:2018pbv}%
  \BibitemOpen
  \bibfield  {author} {\bibinfo {author} {\bibfnamefont {C.~M.}\ \bibnamefont
  {Bender}}, \bibinfo {author} {\bibfnamefont {N.}~\bibnamefont {Hassanpour}},
  \bibinfo {author} {\bibfnamefont {S.~P.}\ \bibnamefont {Klevansky}},\ and\
  \bibinfo {author} {\bibfnamefont {S.}~\bibnamefont {Sarkar}},\ }\bibfield
  {title} {\bibinfo {title} {{$PT$-symmetric quantum field theory in $D$
  dimensions}},\ }\href {https://doi.org/10.1103/PhysRevD.98.125003} {\bibfield
   {journal} {\bibinfo  {journal} {Phys. Rev. D}\ }\textbf {\bibinfo {volume}
  {98}},\ \bibinfo {pages} {125003} (\bibinfo {year} {2018})},\ \Eprint
  {https://arxiv.org/abs/1810.12479} {arXiv:1810.12479 [hep-th]} \BibitemShut
  {NoStop}%
\bibitem [{\citenamefont {Felski}\ \emph {et~al.}(2021)\citenamefont {Felski},
  \citenamefont {Bender}, \citenamefont {Klevansky},\ and\ \citenamefont
  {Sarkar}}]{Felski:2021evi}%
  \BibitemOpen
  \bibfield  {author} {\bibinfo {author} {\bibfnamefont {A.}~\bibnamefont
  {Felski}}, \bibinfo {author} {\bibfnamefont {C.~M.}\ \bibnamefont {Bender}},
  \bibinfo {author} {\bibfnamefont {S.~P.}\ \bibnamefont {Klevansky}},\ and\
  \bibinfo {author} {\bibfnamefont {S.}~\bibnamefont {Sarkar}},\ }\bibfield
  {title} {\bibinfo {title} {{Towards perturbative renormalization of
  {\ensuremath{\phi}}2(i{\ensuremath{\phi}}){\ensuremath{\epsilon}} quantum
  field theory}},\ }\href {https://doi.org/10.1103/PhysRevD.104.085011}
  {\bibfield  {journal} {\bibinfo  {journal} {Phys. Rev. D}\ }\textbf {\bibinfo
  {volume} {104}},\ \bibinfo {pages} {085011} (\bibinfo {year} {2021})},\
  \Eprint {https://arxiv.org/abs/2103.07577} {arXiv:2103.07577 [hep-th]}
  \BibitemShut {NoStop}%
\bibitem [{\citenamefont {Branchina}\ \emph {et~al.}(2021)\citenamefont
  {Branchina}, \citenamefont {Chiavetta},\ and\ \citenamefont
  {Contino}}]{Branchina:2021czr}%
  \BibitemOpen
  \bibfield  {author} {\bibinfo {author} {\bibfnamefont {V.}~\bibnamefont
  {Branchina}}, \bibinfo {author} {\bibfnamefont {A.}~\bibnamefont
  {Chiavetta}},\ and\ \bibinfo {author} {\bibfnamefont {F.}~\bibnamefont
  {Contino}},\ }\bibfield  {title} {\bibinfo {title} {{Study of the
  non-Hermitian PT-symmetric
  g{\ensuremath{\phi}}2(i{\ensuremath{\phi}}){\ensuremath{\varepsilon}} theory:
  Analysis of all orders in {\ensuremath{\varepsilon}} and resummations}},\
  }\href {https://doi.org/10.1103/PhysRevD.104.085010} {\bibfield  {journal}
  {\bibinfo  {journal} {Phys. Rev. D}\ }\textbf {\bibinfo {volume} {104}},\
  \bibinfo {pages} {085010} (\bibinfo {year} {2021})},\ \Eprint
  {https://arxiv.org/abs/2104.12702} {arXiv:2104.12702 [hep-th]} \BibitemShut
  {NoStop}%
\bibitem [{\citenamefont {Beygi}\ \emph {et~al.}(2019)\citenamefont {Beygi},
  \citenamefont {Klevansky},\ and\ \citenamefont {Bender}}]{Beygi:2019qab}%
  \BibitemOpen
  \bibfield  {author} {\bibinfo {author} {\bibfnamefont {A.}~\bibnamefont
  {Beygi}}, \bibinfo {author} {\bibfnamefont {S.~P.}\ \bibnamefont
  {Klevansky}},\ and\ \bibinfo {author} {\bibfnamefont {C.~M.}\ \bibnamefont
  {Bender}},\ }\bibfield  {title} {\bibinfo {title} {{Relativistic
  $PT$-symmetric fermionic theories in 1+1 and 3+1 dimensions}},\ }\href
  {https://doi.org/10.1103/PhysRevA.99.062117} {\bibfield  {journal} {\bibinfo
  {journal} {Phys. Rev. A}\ }\textbf {\bibinfo {volume} {99}},\ \bibinfo
  {pages} {062117} (\bibinfo {year} {2019})},\ \Eprint
  {https://arxiv.org/abs/1904.00878} {arXiv:1904.00878 [math-ph]} \BibitemShut
  {NoStop}%
\bibitem [{\citenamefont {Felski}\ \emph {et~al.}(2020)\citenamefont {Felski},
  \citenamefont {Beygi},\ and\ \citenamefont {Klevansky}}]{Felski:2020vrm}%
  \BibitemOpen
  \bibfield  {author} {\bibinfo {author} {\bibfnamefont {A.}~\bibnamefont
  {Felski}}, \bibinfo {author} {\bibfnamefont {A.}~\bibnamefont {Beygi}},\ and\
  \bibinfo {author} {\bibfnamefont {S.~P.}\ \bibnamefont {Klevansky}},\
  }\bibfield  {title} {\bibinfo {title} {{Non-Hermitian extension of the
  Nambu{\textendash}Jona-Lasinio model in 3+1 and 1+1 dimensions}},\ }\href
  {https://doi.org/10.1103/PhysRevD.101.116001} {\bibfield  {journal} {\bibinfo
   {journal} {Phys. Rev. D}\ }\textbf {\bibinfo {volume} {101}},\ \bibinfo
  {pages} {116001} (\bibinfo {year} {2020})},\ \Eprint
  {https://arxiv.org/abs/2004.04011} {arXiv:2004.04011 [hep-ph]} \BibitemShut
  {NoStop}%
\bibitem [{\citenamefont {Mavromatos}\ and\ \citenamefont
  {Soto}(2021)}]{Mavromatos:2020hfy}%
  \BibitemOpen
  \bibfield  {author} {\bibinfo {author} {\bibfnamefont {N.~E.}\ \bibnamefont
  {Mavromatos}}\ and\ \bibinfo {author} {\bibfnamefont {A.}~\bibnamefont
  {Soto}},\ }\bibfield  {title} {\bibinfo {title} {{Dynamical Majorana neutrino
  masses and axions II: Inclusion of anomaly terms and axial background}},\
  }\href {https://doi.org/10.1016/j.nuclphysb.2020.115275} {\bibfield
  {journal} {\bibinfo  {journal} {Nucl. Phys. B}\ }\textbf {\bibinfo {volume}
  {962}},\ \bibinfo {pages} {115275} (\bibinfo {year} {2021})},\ \Eprint
  {https://arxiv.org/abs/2006.13616} {arXiv:2006.13616 [hep-ph]} \BibitemShut
  {NoStop}%
\bibitem [{\citenamefont {Felski}\ and\ \citenamefont
  {Klevansky}(2021)}]{Felski:2021bdg}%
  \BibitemOpen
  \bibfield  {author} {\bibinfo {author} {\bibfnamefont {A.}~\bibnamefont
  {Felski}}\ and\ \bibinfo {author} {\bibfnamefont {S.~P.}\ \bibnamefont
  {Klevansky}},\ }\bibfield  {title} {\bibinfo {title} {{Fermion and meson mass
  generation in non-Hermitian Nambu{\textendash}Jona-Lasinio models}},\ }\href
  {https://doi.org/10.1103/PhysRevD.103.056007} {\bibfield  {journal} {\bibinfo
   {journal} {Phys. Rev. D}\ }\textbf {\bibinfo {volume} {103}},\ \bibinfo
  {pages} {056007} (\bibinfo {year} {2021})},\ \Eprint
  {https://arxiv.org/abs/2102.01491} {arXiv:2102.01491 [hep-ph]} \BibitemShut
  {NoStop}%
\bibitem [{\citenamefont {Mavromatos}\ \emph {et~al.}(2022)\citenamefont
  {Mavromatos}, \citenamefont {Sarkar},\ and\ \citenamefont
  {Soto}}]{Mavromatos:2021hpe}%
  \BibitemOpen
  \bibfield  {author} {\bibinfo {author} {\bibfnamefont {N.~E.}\ \bibnamefont
  {Mavromatos}}, \bibinfo {author} {\bibfnamefont {S.}~\bibnamefont {Sarkar}},\
  and\ \bibinfo {author} {\bibfnamefont {A.}~\bibnamefont {Soto}},\ }\bibfield
  {title} {\bibinfo {title} {{PT symmetric fermionic field theories with
  axions: Renormalization and dynamical mass generation}},\ }\href
  {https://doi.org/10.1103/PhysRevD.106.015009} {\bibfield  {journal} {\bibinfo
   {journal} {Phys. Rev. D}\ }\textbf {\bibinfo {volume} {106}},\ \bibinfo
  {pages} {015009} (\bibinfo {year} {2022})},\ \Eprint
  {https://arxiv.org/abs/2111.05131} {arXiv:2111.05131 [hep-th]} \BibitemShut
  {NoStop}%
\bibitem [{\citenamefont {Mavromatos}\ and\ \citenamefont
  {Sarkar}(2024)}]{Mavromatos:2024ozk}%
  \BibitemOpen
  \bibfield  {author} {\bibinfo {author} {\bibfnamefont {N.~E.}\ \bibnamefont
  {Mavromatos}}\ and\ \bibinfo {author} {\bibfnamefont {S.}~\bibnamefont
  {Sarkar}},\ }\bibfield  {title} {\bibinfo {title} {{Chern-Simons gravity and
  PT symmetry}},\ }\href {https://doi.org/10.1103/PhysRevD.110.045004}
  {\bibfield  {journal} {\bibinfo  {journal} {Phys. Rev. D}\ }\textbf {\bibinfo
  {volume} {110}},\ \bibinfo {pages} {045004} (\bibinfo {year} {2024})},\
  \Eprint {https://arxiv.org/abs/2402.14513} {arXiv:2402.14513 [hep-th]}
  \BibitemShut {NoStop}%
\bibitem [{\citenamefont {Kuntz}(2025)}]{Kuntz:2024rzu}%
  \BibitemOpen
  \bibfield  {author} {\bibinfo {author} {\bibfnamefont {J.}~\bibnamefont
  {Kuntz}},\ }\bibfield  {title} {\bibinfo {title} {{Unitarity through PT
  symmetry in quantum quadratic gravity}},\ }\href
  {https://doi.org/10.1088/1361-6382/adf606} {\bibfield  {journal} {\bibinfo
  {journal} {Class. Quant. Grav.}\ }\textbf {\bibinfo {volume} {42}},\ \bibinfo
  {pages} {175003} (\bibinfo {year} {2025})},\ \Eprint
  {https://arxiv.org/abs/2410.08278} {arXiv:2410.08278 [hep-th]} \BibitemShut
  {NoStop}%
\bibitem [{\citenamefont
  {Romatschke}(2024{\natexlab{a}})}]{Romatschke:2023fax}%
  \BibitemOpen
  \bibfield  {author} {\bibinfo {author} {\bibfnamefont {P.}~\bibnamefont
  {Romatschke}},\ }\bibfield  {title} {\bibinfo {title} {{Negative Coupling
  $\phi^4$ on the Lattice}},\ }\href {https://doi.org/10.22323/1.453.0367}
  {\bibfield  {journal} {\bibinfo  {journal} {PoS}\ }\textbf {\bibinfo {volume}
  {LATTICE2023}},\ \bibinfo {pages} {367} (\bibinfo {year}
  {2024}{\natexlab{a}})},\ \Eprint {https://arxiv.org/abs/2310.03815}
  {arXiv:2310.03815 [hep-lat]} \BibitemShut {NoStop}%
\bibitem [{\citenamefont {Ogilvie}\ \emph {et~al.}(2025)\citenamefont
  {Ogilvie}, \citenamefont {Schindler},\ and\ \citenamefont
  {Schindler}}]{Ogilvie:2024vde}%
  \BibitemOpen
  \bibfield  {author} {\bibinfo {author} {\bibfnamefont {M.~C.}\ \bibnamefont
  {Ogilvie}}, \bibinfo {author} {\bibfnamefont {M.~A.}\ \bibnamefont
  {Schindler}},\ and\ \bibinfo {author} {\bibfnamefont {S.~T.}\ \bibnamefont
  {Schindler}},\ }\bibfield  {title} {\bibinfo {title} {{Exotic phases in
  finite-density {\ensuremath{\mathbb{Z}}}$_{3}$ theories}},\ }\href
  {https://doi.org/10.1007/JHEP03(2025)077} {\bibfield  {journal} {\bibinfo
  {journal} {JHEP}\ }\textbf {\bibinfo {volume} {03}},\ \bibinfo {pages}
  {077}},\ \Eprint {https://arxiv.org/abs/2411.11773} {arXiv:2411.11773
  [hep-ph]} \BibitemShut {NoStop}%
\bibitem [{\citenamefont {Ai}\ \emph {et~al.}(2022)\citenamefont {Ai},
  \citenamefont {Bender},\ and\ \citenamefont {Sarkar}}]{Ai:2022csx}%
  \BibitemOpen
  \bibfield  {author} {\bibinfo {author} {\bibfnamefont {W.-Y.}\ \bibnamefont
  {Ai}}, \bibinfo {author} {\bibfnamefont {C.~M.}\ \bibnamefont {Bender}},\
  and\ \bibinfo {author} {\bibfnamefont {S.}~\bibnamefont {Sarkar}},\
  }\bibfield  {title} {\bibinfo {title} {{PT-symmetric
  -g{\ensuremath{\varphi}}4 theory}},\ }\href
  {https://doi.org/10.1103/PhysRevD.106.125016} {\bibfield  {journal} {\bibinfo
   {journal} {Phys. Rev. D}\ }\textbf {\bibinfo {volume} {106}},\ \bibinfo
  {pages} {125016} (\bibinfo {year} {2022})},\ \Eprint
  {https://arxiv.org/abs/2209.07897} {arXiv:2209.07897 [hep-th]} \BibitemShut
  {NoStop}%
\bibitem [{\citenamefont
  {Romatschke}(2023{\natexlab{a}})}]{Romatschke:2022jqg}%
  \BibitemOpen
  \bibfield  {author} {\bibinfo {author} {\bibfnamefont {P.}~\bibnamefont
  {Romatschke}},\ }\bibfield  {title} {\bibinfo {title} {{A solvable quantum
  field theory with asymptotic freedom in (3+1) dimensions}},\ }\href
  {https://doi.org/10.1142/S0217751X23501579} {\bibfield  {journal} {\bibinfo
  {journal} {Int. J. Mod. Phys. A}\ }\textbf {\bibinfo {volume} {38}},\
  \bibinfo {pages} {2350157} (\bibinfo {year} {2023}{\natexlab{a}})},\ \Eprint
  {https://arxiv.org/abs/2211.15683} {arXiv:2211.15683 [hep-th]} \BibitemShut
  {NoStop}%
\bibitem [{\citenamefont {Lawrence}\ \emph {et~al.}(2023)\citenamefont
  {Lawrence}, \citenamefont {Weller}, \citenamefont {Peterson},\ and\
  \citenamefont {Romatschke}}]{Lawrence:2023woz}%
  \BibitemOpen
  \bibfield  {author} {\bibinfo {author} {\bibfnamefont {S.}~\bibnamefont
  {Lawrence}}, \bibinfo {author} {\bibfnamefont {R.}~\bibnamefont {Weller}},
  \bibinfo {author} {\bibfnamefont {C.}~\bibnamefont {Peterson}},\ and\
  \bibinfo {author} {\bibfnamefont {P.}~\bibnamefont {Romatschke}},\ }\bibfield
   {title} {\bibinfo {title} {{Instantons, analytic continuation, and
  PT-symmetric field theory}},\ }\href
  {https://doi.org/10.1103/PhysRevD.108.085013} {\bibfield  {journal} {\bibinfo
   {journal} {Phys. Rev. D}\ }\textbf {\bibinfo {volume} {108}},\ \bibinfo
  {pages} {085013} (\bibinfo {year} {2023})},\ \Eprint
  {https://arxiv.org/abs/2303.01470} {arXiv:2303.01470 [hep-th]} \BibitemShut
  {NoStop}%
\bibitem [{\citenamefont {Weller}(2023)}]{Weller:2023jhc}%
  \BibitemOpen
  \bibfield  {author} {\bibinfo {author} {\bibfnamefont {R.~D.}\ \bibnamefont
  {Weller}},\ }\bibfield  {title} {\bibinfo {title} {{Can negative bare
  couplings make sense? The $\vec{\phi}^4$ theory at large $N$}},\ }\href@noop
  {} {\  (\bibinfo {year} {2023})},\ \Eprint {https://arxiv.org/abs/2310.02516}
  {arXiv:2310.02516 [hep-th]} \BibitemShut {NoStop}%
\bibitem [{\citenamefont {Chen}\ and\ \citenamefont
  {Sarkar}(2025)}]{Chen:2024ynx}%
  \BibitemOpen
  \bibfield  {author} {\bibinfo {author} {\bibfnamefont {L.}~\bibnamefont
  {Chen}}\ and\ \bibinfo {author} {\bibfnamefont {S.}~\bibnamefont {Sarkar}},\
  }\bibfield  {title} {\bibinfo {title} {{Phases of quartic scalar theories and
  PT symmetry}},\ }\href {https://doi.org/10.1103/PhysRevD.111.025001}
  {\bibfield  {journal} {\bibinfo  {journal} {Phys. Rev. D}\ }\textbf {\bibinfo
  {volume} {111}},\ \bibinfo {pages} {025001} (\bibinfo {year} {2025})},\
  \Eprint {https://arxiv.org/abs/2409.05439} {arXiv:2409.05439 [quant-ph]}
  \BibitemShut {NoStop}%
\bibitem [{\citenamefont {Romatschke}(2025)}]{Romatschke:2024cld}%
  \BibitemOpen
  \bibfield  {author} {\bibinfo {author} {\bibfnamefont {P.}~\bibnamefont
  {Romatschke}},\ }\bibfield  {title} {\bibinfo {title} {{On the negative
  coupling O(N) model in 2d at high temperature}},\ }\href
  {https://doi.org/10.1007/JHEP04(2025)012} {\bibfield  {journal} {\bibinfo
  {journal} {JHEP}\ }\textbf {\bibinfo {volume} {04}},\ \bibinfo {pages}
  {012}},\ \Eprint {https://arxiv.org/abs/2412.10496} {arXiv:2412.10496
  [hep-th]} \BibitemShut {NoStop}%
\bibitem [{\citenamefont {Barberena}(2025)}]{Barberena:2025ibo}%
  \BibitemOpen
  \bibfield  {author} {\bibinfo {author} {\bibfnamefont {D.}~\bibnamefont
  {Barberena}},\ }\bibfield  {title} {\bibinfo {title} {{Large N vector models
  in the Hamiltonian framework}},\ }\href@noop {} {\  (\bibinfo {year}
  {2025})},\ \Eprint {https://arxiv.org/abs/2502.08031} {arXiv:2502.08031
  [hep-th]} \BibitemShut {NoStop}%
\bibitem [{\citenamefont {Fring}\ and\ \citenamefont
  {Taira}(2020)}]{Fring:2019hue}%
  \BibitemOpen
  \bibfield  {author} {\bibinfo {author} {\bibfnamefont {A.}~\bibnamefont
  {Fring}}\ and\ \bibinfo {author} {\bibfnamefont {T.}~\bibnamefont {Taira}},\
  }\bibfield  {title} {\bibinfo {title} {{Goldstone bosons in different
  PT-regimes of non-Hermitian scalar quantum field theories}},\ }\href
  {https://doi.org/10.1016/j.nuclphysb.2019.114834} {\bibfield  {journal}
  {\bibinfo  {journal} {Nucl. Phys. B}\ }\textbf {\bibinfo {volume} {950}},\
  \bibinfo {pages} {114834} (\bibinfo {year} {2020})},\ \Eprint
  {https://arxiv.org/abs/1906.05738} {arXiv:1906.05738 [hep-th]} \BibitemShut
  {NoStop}%
\bibitem [{\citenamefont {Fring}\ and\ \citenamefont
  {Taira}(2022)}]{Fring:2020bvr}%
  \BibitemOpen
  \bibfield  {author} {\bibinfo {author} {\bibfnamefont {A.}~\bibnamefont
  {Fring}}\ and\ \bibinfo {author} {\bibfnamefont {T.}~\bibnamefont {Taira}},\
  }\bibfield  {title} {\bibinfo {title} {{Massive gauge particles versus
  Goldstone bosons in non-Hermitian non-Abelian gauge theory}},\ }\href
  {https://doi.org/10.1140/epjp/s13360-022-02889-z} {\bibfield  {journal}
  {\bibinfo  {journal} {Eur. Phys. J. Plus}\ }\textbf {\bibinfo {volume}
  {137}},\ \bibinfo {pages} {716} (\bibinfo {year} {2022})},\ \Eprint
  {https://arxiv.org/abs/2004.00723} {arXiv:2004.00723 [hep-th]} \BibitemShut
  {NoStop}%
\bibitem [{\citenamefont {Romatschke}\ \emph {et~al.}(2024)\citenamefont
  {Romatschke}, \citenamefont {Su},\ and\ \citenamefont
  {Weller}}]{Romatschke:2024hpb}%
  \BibitemOpen
  \bibfield  {author} {\bibinfo {author} {\bibfnamefont {P.}~\bibnamefont
  {Romatschke}}, \bibinfo {author} {\bibfnamefont {C.-W.}\ \bibnamefont {Su}},\
  and\ \bibinfo {author} {\bibfnamefont {R.}~\bibnamefont {Weller}},\
  }\bibfield  {title} {\bibinfo {title} {{Mass generation in an Abelian gauge
  theory with multiple scalar fields and no tree-level dimensionful
  couplings}},\ }\href {https://doi.org/10.1103/PhysRevD.110.113006} {\bibfield
   {journal} {\bibinfo  {journal} {Phys. Rev. D}\ }\textbf {\bibinfo {volume}
  {110}},\ \bibinfo {pages} {113006} (\bibinfo {year} {2024})},\ \Eprint
  {https://arxiv.org/abs/2405.00088} {arXiv:2405.00088 [hep-ph]} \BibitemShut
  {NoStop}%
\bibitem [{\citenamefont {Abbott}\ \emph {et~al.}(1976)\citenamefont {Abbott},
  \citenamefont {Kang},\ and\ \citenamefont {Schnitzer}}]{Abbott:1975bn}%
  \BibitemOpen
  \bibfield  {author} {\bibinfo {author} {\bibfnamefont {L.~F.}\ \bibnamefont
  {Abbott}}, \bibinfo {author} {\bibfnamefont {J.~S.}\ \bibnamefont {Kang}},\
  and\ \bibinfo {author} {\bibfnamefont {H.~J.}\ \bibnamefont {Schnitzer}},\
  }\bibfield  {title} {\bibinfo {title} {{Bound States, Tachyons, and
  Restoration of Symmetry in the 1/N Expansion}},\ }\href
  {https://doi.org/10.1103/PhysRevD.13.2212} {\bibfield  {journal} {\bibinfo
  {journal} {Phys. Rev. D}\ }\textbf {\bibinfo {volume} {13}},\ \bibinfo
  {pages} {2212} (\bibinfo {year} {1976})}\BibitemShut {NoStop}%
\bibitem [{\citenamefont {Linde}(1977)}]{Linde:1976qh}%
  \BibitemOpen
  \bibfield  {author} {\bibinfo {author} {\bibfnamefont {A.~D.}\ \bibnamefont
  {Linde}},\ }\bibfield  {title} {\bibinfo {title} {{1/n-Expansion, Vacuum
  Stability and Quark Confinement}},\ }\href
  {https://doi.org/10.1016/0550-3213(77)90112-2} {\bibfield  {journal}
  {\bibinfo  {journal} {Nucl. Phys. B}\ }\textbf {\bibinfo {volume} {125}},\
  \bibinfo {pages} {369} (\bibinfo {year} {1977})}\BibitemShut {NoStop}%
\bibitem [{\citenamefont
  {Romatschke}(2023{\natexlab{b}})}]{Romatschke:2023sce}%
  \BibitemOpen
  \bibfield  {author} {\bibinfo {author} {\bibfnamefont {P.}~\bibnamefont
  {Romatschke}},\ }\bibfield  {title} {\bibinfo {title} {{What if
  {\ensuremath{\phi}}4 theory in 4 dimensions is non-trivial in the
  continuum?}},\ }\href {https://doi.org/10.1016/j.physletb.2023.138270}
  {\bibfield  {journal} {\bibinfo  {journal} {Phys. Lett. B}\ }\textbf
  {\bibinfo {volume} {847}},\ \bibinfo {pages} {138270} (\bibinfo {year}
  {2023}{\natexlab{b}})},\ \Eprint {https://arxiv.org/abs/2305.05678}
  {arXiv:2305.05678 [hep-th]} \BibitemShut {NoStop}%
\bibitem [{\citenamefont
  {Romatschke}(2024{\natexlab{b}})}]{Romatschke:2023ztk}%
  \BibitemOpen
  \bibfield  {author} {\bibinfo {author} {\bibfnamefont {P.}~\bibnamefont
  {Romatschke}},\ }\bibfield  {title} {\bibinfo {title} {{Quantum Field Theory
  in Large-$N$ Wonderland: Three Lectures}},\ }\href
  {https://doi.org/10.5506/APhysPolB.55.4-A2} {\bibfield  {journal} {\bibinfo
  {journal} {Acta Phys. Polon. B}\ }\textbf {\bibinfo {volume} {55}},\ \bibinfo
  {pages} {4} (\bibinfo {year} {2024}{\natexlab{b}})},\ \Eprint
  {https://arxiv.org/abs/2310.00048} {arXiv:2310.00048 [hep-th]} \BibitemShut
  {NoStop}%
\bibitem [{\citenamefont
  {Romatschke}(2024{\natexlab{c}})}]{Romatschke:2024yhx}%
  \BibitemOpen
  \bibfield  {author} {\bibinfo {author} {\bibfnamefont {P.}~\bibnamefont
  {Romatschke}},\ }\bibfield  {title} {\bibinfo {title} {{Alternative to
  perturbative renormalization in (3+1)-dimensional field theories}},\ }\href
  {https://doi.org/10.1103/PhysRevD.109.116020} {\bibfield  {journal} {\bibinfo
   {journal} {Phys. Rev. D}\ }\textbf {\bibinfo {volume} {109}},\ \bibinfo
  {pages} {116020} (\bibinfo {year} {2024}{\natexlab{c}})},\ \Eprint
  {https://arxiv.org/abs/2401.06847} {arXiv:2401.06847 [hep-th]} \BibitemShut
  {NoStop}%
\bibitem [{\citenamefont {Lawrence}\ \emph {et~al.}(2022)\citenamefont
  {Lawrence}, \citenamefont {Oh},\ and\ \citenamefont
  {Yamauchi}}]{Lawrence:2022afv}%
  \BibitemOpen
  \bibfield  {author} {\bibinfo {author} {\bibfnamefont {S.}~\bibnamefont
  {Lawrence}}, \bibinfo {author} {\bibfnamefont {H.}~\bibnamefont {Oh}},\ and\
  \bibinfo {author} {\bibfnamefont {Y.}~\bibnamefont {Yamauchi}},\ }\bibfield
  {title} {\bibinfo {title} {{Lattice scalar field theory at complex
  coupling}},\ }\href {https://doi.org/10.1103/PhysRevD.106.114503} {\bibfield
  {journal} {\bibinfo  {journal} {Phys. Rev. D}\ }\textbf {\bibinfo {volume}
  {106}},\ \bibinfo {pages} {114503} (\bibinfo {year} {2022})},\ \Eprint
  {https://arxiv.org/abs/2205.12303} {arXiv:2205.12303 [hep-lat]} \BibitemShut
  {NoStop}%
\bibitem [{\citenamefont {Romatschke}(2026)}]{Romatschke:2026tam}%
  \BibitemOpen
  \bibfield  {author} {\bibinfo {author} {\bibfnamefont {P.}~\bibnamefont
  {Romatschke}},\ }\bibfield  {title} {\bibinfo {title} {{On self-dualities for
  scalar $\phi^4$ theory}},\ }\href@noop {} {\  (\bibinfo {year} {2026})},\
  \Eprint {https://arxiv.org/abs/2602.18286} {arXiv:2602.18286 [hep-th]}
  \BibitemShut {NoStop}%
\bibitem [{\citenamefont {Jones}\ and\ \citenamefont
  {Mateo}(2006)}]{Jones:2006qs}%
  \BibitemOpen
  \bibfield  {author} {\bibinfo {author} {\bibfnamefont {H.~F.}\ \bibnamefont
  {Jones}}\ and\ \bibinfo {author} {\bibfnamefont {J.}~\bibnamefont {Mateo}},\
  }\bibfield  {title} {\bibinfo {title} {{An Equivalent Hermitian Hamiltonian
  for the non-Hermitian -x**4 potential}},\ }\href
  {https://doi.org/10.1103/PhysRevD.73.085002} {\bibfield  {journal} {\bibinfo
  {journal} {Phys. Rev. D}\ }\textbf {\bibinfo {volume} {73}},\ \bibinfo
  {pages} {085002} (\bibinfo {year} {2006})},\ \Eprint
  {https://arxiv.org/abs/quant-ph/0601188} {arXiv:quant-ph/0601188}
  \BibitemShut {NoStop}%
\bibitem [{\citenamefont {Laine}\ and\ \citenamefont
  {Vuorinen}(2016)}]{Laine:2016hma}%
  \BibitemOpen
  \bibfield  {author} {\bibinfo {author} {\bibfnamefont {M.}~\bibnamefont
  {Laine}}\ and\ \bibinfo {author} {\bibfnamefont {A.}~\bibnamefont
  {Vuorinen}},\ }\href {https://doi.org/10.1007/978-3-319-31933-9} {\emph
  {\bibinfo {title} {{Basics of Thermal Field Theory}}}},\ Vol.\ \bibinfo
  {volume} {925}\ (\bibinfo  {publisher} {Springer},\ \bibinfo {year} {2016})\
  \Eprint {https://arxiv.org/abs/1701.01554} {arXiv:1701.01554 [hep-ph]}
  \BibitemShut {NoStop}%
\bibitem [{\citenamefont {Romatschke}(2019)}]{Romatschke:2019rjk}%
  \BibitemOpen
  \bibfield  {author} {\bibinfo {author} {\bibfnamefont {P.}~\bibnamefont
  {Romatschke}},\ }\bibfield  {title} {\bibinfo {title} {{Simple
  non-perturbative resummation schemes beyond mean-field: case study for scalar
  $\phi^4$ theory in 1+1 dimensions}},\ }\href
  {https://doi.org/10.1007/JHEP03(2019)149} {\bibfield  {journal} {\bibinfo
  {journal} {JHEP}\ }\textbf {\bibinfo {volume} {03}},\ \bibinfo {pages}
  {149}},\ \Eprint {https://arxiv.org/abs/1901.05483} {arXiv:1901.05483
  [hep-th]} \BibitemShut {NoStop}%
\bibitem [{\citenamefont {Kamata}(2024{\natexlab{a}})}]{Kamata:2023opn}%
  \BibitemOpen
  \bibfield  {author} {\bibinfo {author} {\bibfnamefont {S.}~\bibnamefont
  {Kamata}},\ }\bibfield  {title} {\bibinfo {title} {{Exact WKB analysis for
  PT-symmetric quantum mechanics: Study of the Ai-Bender-Sarkar conjecture}},\
  }\href {https://doi.org/10.1103/PhysRevD.109.085023} {\bibfield  {journal}
  {\bibinfo  {journal} {Phys. Rev. D}\ }\textbf {\bibinfo {volume} {109}},\
  \bibinfo {pages} {085023} (\bibinfo {year} {2024}{\natexlab{a}})},\ \Eprint
  {https://arxiv.org/abs/2401.00574} {arXiv:2401.00574 [hep-th]} \BibitemShut
  {NoStop}%
\bibitem [{\citenamefont {Kamata}(2024{\natexlab{b}})}]{Kamata:2024tyb}%
  \BibitemOpen
  \bibfield  {author} {\bibinfo {author} {\bibfnamefont {S.}~\bibnamefont
  {Kamata}},\ }\bibfield  {title} {\bibinfo {title} {{Exact quantization
  conditions and full transseries structures for PT symmetric anharmonic
  oscillators}},\ }\href {https://doi.org/10.1103/PhysRevD.110.045022}
  {\bibfield  {journal} {\bibinfo  {journal} {Phys. Rev. D}\ }\textbf {\bibinfo
  {volume} {110}},\ \bibinfo {pages} {045022} (\bibinfo {year}
  {2024}{\natexlab{b}})},\ \Eprint {https://arxiv.org/abs/2406.01230}
  {arXiv:2406.01230 [hep-th]} \BibitemShut {NoStop}%
\bibitem [{\citenamefont {Schaich}\ and\ \citenamefont
  {Loinaz}(2009)}]{Schaich:2009jk}%
  \BibitemOpen
  \bibfield  {author} {\bibinfo {author} {\bibfnamefont {D.}~\bibnamefont
  {Schaich}}\ and\ \bibinfo {author} {\bibfnamefont {W.}~\bibnamefont
  {Loinaz}},\ }\bibfield  {title} {\bibinfo {title} {{An Improved lattice
  measurement of the critical coupling in phi(2)**4 theory}},\ }\href
  {https://doi.org/10.1103/PhysRevD.79.056008} {\bibfield  {journal} {\bibinfo
  {journal} {Phys. Rev. D}\ }\textbf {\bibinfo {volume} {79}},\ \bibinfo
  {pages} {056008} (\bibinfo {year} {2009})},\ \Eprint
  {https://arxiv.org/abs/0902.0045} {arXiv:0902.0045 [hep-lat]} \BibitemShut
  {NoStop}%
\bibitem [{\citenamefont {Aizenman}\ and\ \citenamefont
  {Duminil-Copin}(2021)}]{Aizenman:2019yuo}%
  \BibitemOpen
  \bibfield  {author} {\bibinfo {author} {\bibfnamefont {M.}~\bibnamefont
  {Aizenman}}\ and\ \bibinfo {author} {\bibfnamefont {H.}~\bibnamefont
  {Duminil-Copin}},\ }\bibfield  {title} {\bibinfo {title} {{Marginal
  triviality of the scaling limits of critical 4D Ising and $\phi_4^4$
  models}},\ }\href {https://doi.org/10.4007/annals.2021.194.1.3} {\bibfield
  {journal} {\bibinfo  {journal} {Annals Math.}\ }\textbf {\bibinfo {volume}
  {194}},\ \bibinfo {pages} {163} (\bibinfo {year} {2021})},\ \Eprint
  {https://arxiv.org/abs/1912.07973} {arXiv:1912.07973 [math-ph]} \BibitemShut
  {NoStop}%
\bibitem [{\citenamefont
  {Romatschke}(2024{\natexlab{d}})}]{Romatschke:2023ogd}%
  \BibitemOpen
  \bibfield  {author} {\bibinfo {author} {\bibfnamefont {P.}~\bibnamefont
  {Romatschke}},\ }\bibfield  {title} {\bibinfo {title} {{A loophole in the
  proofs of asymptotic freedom and quantum triviality}},\ }\href
  {https://doi.org/10.22323/1.449.0500} {\bibfield  {journal} {\bibinfo
  {journal} {PoS}\ }\textbf {\bibinfo {volume} {EPS-HEP2023}},\ \bibinfo
  {pages} {500} (\bibinfo {year} {2024}{\natexlab{d}})},\ \Eprint
  {https://arxiv.org/abs/2310.18414} {arXiv:2310.18414 [hep-th]} \BibitemShut
  {NoStop}%
\bibitem [{\citenamefont {{Aizenman}}\ and\ \citenamefont
  {{Graham}}(1983)}]{1983NuPhB.225..261A}%
  \BibitemOpen
  \bibfield  {author} {\bibinfo {author} {\bibfnamefont {M.}~\bibnamefont
  {{Aizenman}}}\ and\ \bibinfo {author} {\bibfnamefont {R.}~\bibnamefont
  {{Graham}}},\ }\bibfield  {title} {\bibinfo {title} {{On the renormalized
  coupling constant and the susceptibility in {\ensuremath{\varphi}}
  $_{4}$$^{4}$ field theory and the Ising model in four dimensions}},\ }\href
  {https://doi.org/10.1016/0550-3213(83)90053-6} {\bibfield  {journal}
  {\bibinfo  {journal} {Nuclear Physics B}\ }\textbf {\bibinfo {volume}
  {225}},\ \bibinfo {pages} {261} (\bibinfo {year} {1983})}\BibitemShut
  {NoStop}%
\bibitem [{\citenamefont {Fröhlich}(1982)}]{Frohlich:1982tw}%
  \BibitemOpen
  \bibfield  {author} {\bibinfo {author} {\bibfnamefont {J.}~\bibnamefont
  {Fröhlich}},\ }\bibfield  {title} {\bibinfo {title} {{On the Triviality of
  Lambda (phi**4) in D-Dimensions Theories and the Approach to the Critical
  Point in D$\geq$ Four-Dimensions}},\ }\href
  {https://doi.org/10.1016/0550-3213(82)90088-8} {\bibfield  {journal}
  {\bibinfo  {journal} {Nucl. Phys. B}\ }\textbf {\bibinfo {volume} {200}},\
  \bibinfo {pages} {281} (\bibinfo {year} {1982})}\BibitemShut {NoStop}%
\bibitem [{\citenamefont {Wilson}(1983)}]{Wilson:1983xri}%
  \BibitemOpen
  \bibfield  {author} {\bibinfo {author} {\bibfnamefont {K.~G.}\ \bibnamefont
  {Wilson}},\ }\bibfield  {title} {\bibinfo {title} {{The renormalization group
  and critical phenomena}},\ }\href {https://doi.org/10.1103/RevModPhys.55.583}
  {\bibfield  {journal} {\bibinfo  {journal} {Rev. Mod. Phys.}\ }\textbf
  {\bibinfo {volume} {55}},\ \bibinfo {pages} {583} (\bibinfo {year}
  {1983})}\BibitemShut {NoStop}%
\end{thebibliography}%
\end{document}